\documentclass[fleqn,usenatbib]{mnras}

\usepackage{mathptmx}

\usepackage[T1]{fontenc}
\usepackage{ae,aecompl}

\hypersetup{draft}

\usepackage{soul}

\usepackage{graphicx}	
\usepackage{amsmath}	
\usepackage{amssymb}	

\usepackage{color}

\title{A 9-Hr CV With One Outburst in 4 Years of {\em Kepler} Data}

\author[Yu et al.]{
Zhifei Yu$^1$,
J.R.~Thorstensen$^2$, 
S.~Rappaport$^3$,   
A.~Mann$^4$,
T.~Jacobs$^5$,
\newauthor
L.~Nelson$^6$, 
B.T.~G\"ansicke$^7$,
D.~LaCourse$^8$,
T.~Borkovits$^{9,10}$,
J.~Aiken$^{6}$,
\newauthor
D.~Steeghs$^{7}$,
O.~Toloza$^{7}$,
A.~Vanderburg$^{12,13}$, 
and D.\,N.\,C.~Lin$^{14}$
\\
$^1$ Phillips Academy, 180 Main St., Andover, MA 01810; zyu20@andover.edu \\
$^{2}$ Department of Physics and Astronomy, Dartmouth College, 239 Wilder Hall, Hanover, NH 03755; john.r.thorstensen@dartmouth.edu \\
$^{3}$ Department of Physics, and Kavli Institute for Astrophysics and Space Research, M.I.T., Cambridge, MA 02139, USA; sar@mit.edu \\
$^{4}$ Department of Physics and Astronomy, University of North Carolina at Chapel Hill, Chapel Hill, NC 27599-3255, USA \\
$^{5}$ Amateur Astronomer, 12812 SE 69th Place Bellevue, WA 98006 \\
$^{6}$ Department of Physics and Astronomy, Bishop's University, 2600 College St., Sherbrooke, QC J1M 1Z7; lnelson@ubishops.ca \\  
$^7$ Department of Physics, University of Warwick, Coventry CV4 7AL, UK \\ 
$^{8}$ Amateur Astronomer, 7507 52nd Place NE Marysville, WA 98270 \\
$^9$ Baja Astronomical Observatory of Szeged University, H-6500 Baja, Szegedi \'{u}t, Kt. 766, Hungary; borko@electra.bajaobs.hu \\
$^{10}$ Konkoly Observatory, Research Centre for Astronomy and Earth Sciences, Hungarian Academy of Sciences, \\
$^{11}$  Astronomy and Astrophysics Group, University of Warwick, Coventry, CV4 7AL UK \\
$^{12}$ Department of Astronomy, The University of Texas at Austin, 2515 Speedway, Stop C1400, Austin, TX 78712 \\
$^{13}$ NASA Sagan Fellow \\
$^{14}$ Department of Astronomy and Astrophysics, University of California, Santa Cruz, CA 95064, USA \\
}

\date{Submitted 2017 February 28}

\pubyear{2017}

\begin{document}
\label{firstpage}
\pagerange{\pageref{firstpage}--\pageref{lastpage}}
\maketitle

\begin{abstract}
During a visual search through the {\em Kepler} main-field lightcurves, we have discovered a cataclysmic variable (CV) that experienced only a single 4-day long outburst over four years, rising to three times the quiescent flux. During the four years of non-outburst data the {\em Kepler} photometry of KIC 5608384 exhibits ellipsoidal light variations (`ELV') with a $\sim$12\% amplitude and period of 8.7 hours.  Follow-up ground-based spectral observations have yielded a high-quality radial velocity curve and the associated mass function.  Additionally, H$\alpha$ emission lines were present in the spectra even though these were  taken while the source was presumably in quiescence.  These emission lines are at least partially eclipsed by the companion K star.  We utilize the available constraints of the mass function, the ELV amplitude, Roche-lobe filling condition, and inferred radius of the K star to derive the system masses and orbital inclination angle: $M_{\rm wd} \simeq 0.46 \pm 0.02 \, M_\odot$, $M_{\rm K} \simeq 0.41 \pm 0.03 \, M_\odot$, and $i \gtrsim 70^\circ$. The value of $M_{\rm wd}$ is the lowest reported for any accreting WD in a cataclysmic variable.  We have also run binary evolution models using {\tt MESA} to infer the most likely parameters of the pre-cataclysmic binary.  Using the mass-transfer rates from the model evolution tracks we conclude that although the rates are close to the critical value for accretion disk stability, we expect KIC 5608384 to exhibit dwarf nova outbursts.  We also conclude that the accreting white dwarf most likely descended from a hot subdwarf and, most notably, that this binary is one of the first {\sl bona fide} examples of a progenitor of AM CVn binaries to have evolved through the CV channel.
\end{abstract}




\section{Introduction} 
\label{sec:intro}

Cataclysmic variables (CVs) are close binaries in which a white dwarf accretes from a Roche-lobe filling companion star (see, e.g., \citealt{warner95}). Angular momentum losses drive the evolution of nearly all these systems from long to short orbital periods, with the mass of the donor monotonically decreasing. Observationally, CVs are an extremely heterogeneous class, with manifold characteristics in both their long-term variability and spectroscopic appearances. Most CVs contain weakly or non-magnetic white dwarfs, and the material lost by the donor star forms an accretion disc around the white dwarf. Depending on the mass transfer rate, and the physical dimensions of the CV, these discs undergo quasi-periodic thermal instabilities, called dwarf nova outbursts \citep{meyer81} with recurrence times of weeks to decades, or persist in a quasi-steady high-luminosity state. The vast majority of known CVs were identified thanks to these outbursts. While it is clear that the resulting CV sample is observationally biased, the selection effects are extremely difficult to quantify beyond the simplistic statement that CVs with infrequent outbursts, or low-amplitude outbursts, are very likely underrepresented \citep{breedt14}. 

Historically, it was assumed that most CVs evolved as a result of stable Roche-lobe overflow from a donor that is somewhat less massive than its white dwarf companion; this mass loss causes the donor to be driven slightly out of its thermal equilibrium configuration. Consequently, the physical properties of CV donor stars should resemble main-sequence stars of the same mass. With $\simeq90$ per cent of the $\simeq1400$  CVs with known orbital periods \citep{ritter03} having $80\,\mathrm{min}<P_\mathrm{orb}<10$\,h, the donors are hence expected to gradually morph through the spectral types G, K, and M until they eventually become brown dwarfs that are so dim that they are very difficult to detect. 

However, a number of CVs have been found that defy that prediction: \citet{augusteijn86} and \citet{thorstensen02_eipsc} identified two dwarf novae with periods around one hour, significantly shorter than the predicted minimum period for CVs \citep{paczynski81, rappaport82}, but containing over-luminous donors that resemble K-type stars. These systems very likely initiated thermal timescale mass transfer from much more massive donors, $\ga1.2-1.5\,M_\odot$, and are now accreting from the highly evolved, stripped cores, and are therefore much hotter than the donors in normal CVs \citep{augusteijn86, thorstensen02_eipsc}.  A small number of additional systems that have undergone thermal-time scale mass transfer have been discovered, mostly serendipitously \citep{thorstensen02_qzser, rodriguez-gil09, rebassa14, thorstensen15, harrison18}, thus demonstrating that they represent a non-negligible contribution to the overall CV population.  This evolutionary channel is thought to contribute to the formation of the ultra-short period AM\,CVn stars \citep{podsiadlowski03}, which are strong low-frequency gravitational wave sources that are expected to be detectable by \textit{LISA}.  For a review of these and related CV evolutionary channels see \citet{goliasch15} and \citet{kalomeni16}.

In this paper, we report on a new CV with a significantly evolved donor star, discovered from a single outburst detected in its four year-long \textit{Kepler} light curve~--~underscoring the fact that selection biases may result in a under-representation of these systems among the known CV population.  In Sect.~\ref{sec:Kepler} we describe the discovery and the {\em Kepler} observations of KIC 5608384 in the {\em Kepler} main field.  This CV is quite unusual in that it experienced only a single outburst over four years of observations.  We utilize a series of ground-based spectra of KIC 5608384 in Sect.~\ref{sec:spectra} to derive the radial velocity curve for the companion K star, from which we derive the mass function.  We utilize {\em Swift} and {\em GALEX} UV flux measurements to constrain the white dwarf temperature (see Sect.~\ref{sec:UV}). In Sect.~\ref{sec:SED}, we use the spectral energy distribution (`SED') coupled with the Gaia distance to infer the radius of the K-star companion.  Estimates for the long-term accretion luminosity of the system and the corresponding mass transfer rate, based on $T_{\rm eff}$ of the white-dwarf,  are given in Sect.~\ref{sec:lum}.  The system parameters are deduced in Sect.~\ref{sec:mcmc} by utilizing the various constraints on the system via an MCMC analysis. In Sect.~\ref{sec:origin} we discuss the origin of KIC 5608384, including a likely thermal-timescale mass-transfer phase.  We use binary evolution tracks computed with {\tt MESA} to model the current status of KIC 5608384 as well as to understand its future, which appears to include a phase as an AM CVn system.  We summarize our results in Sect.~\ref{sec:concl}.

\section{{\em Kepler} Observations} 
\label{sec:Kepler}

\begin{figure}
\begin{center}
\includegraphics[width=1.01 \columnwidth]{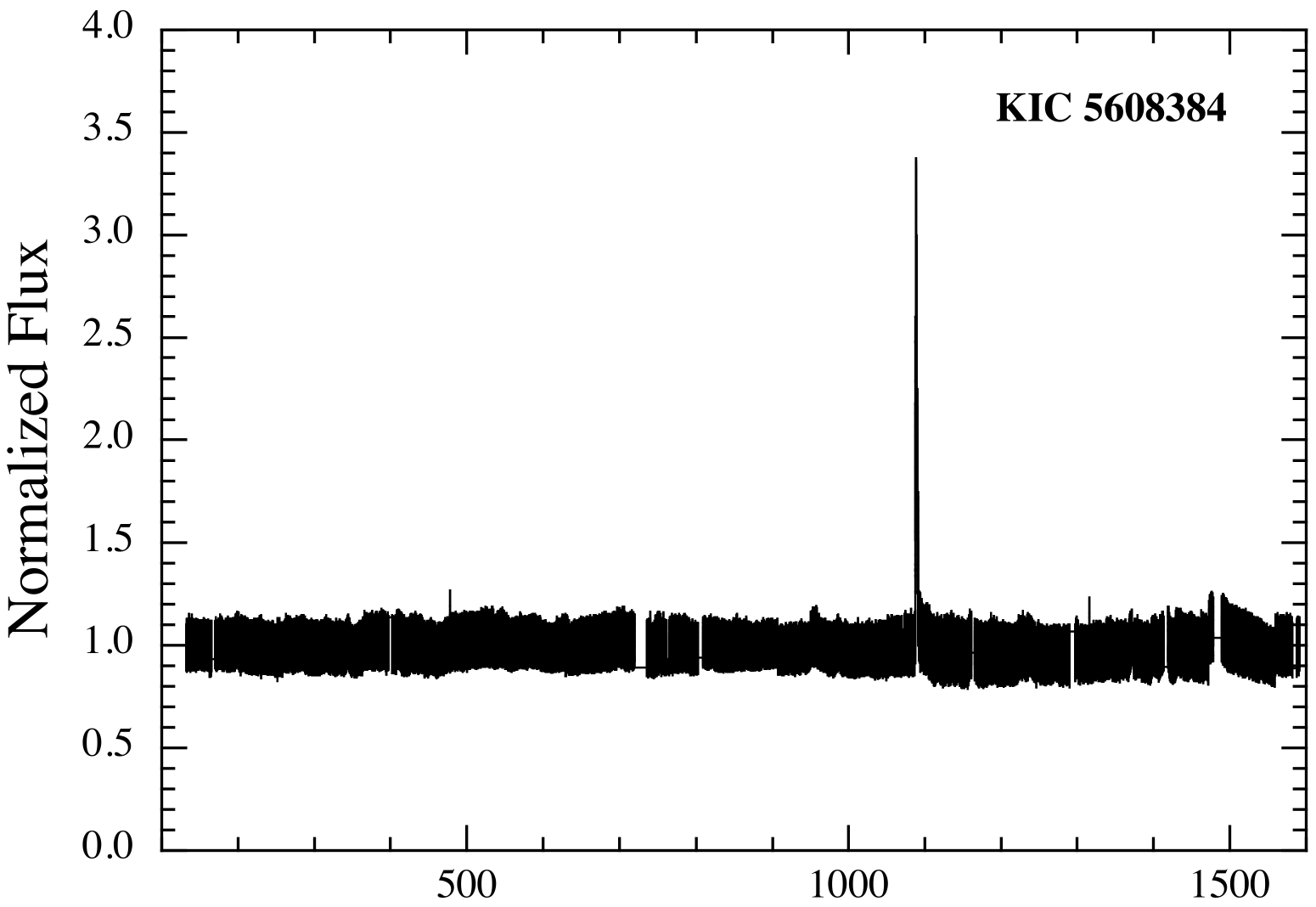}
\includegraphics[width=1.01 \columnwidth]{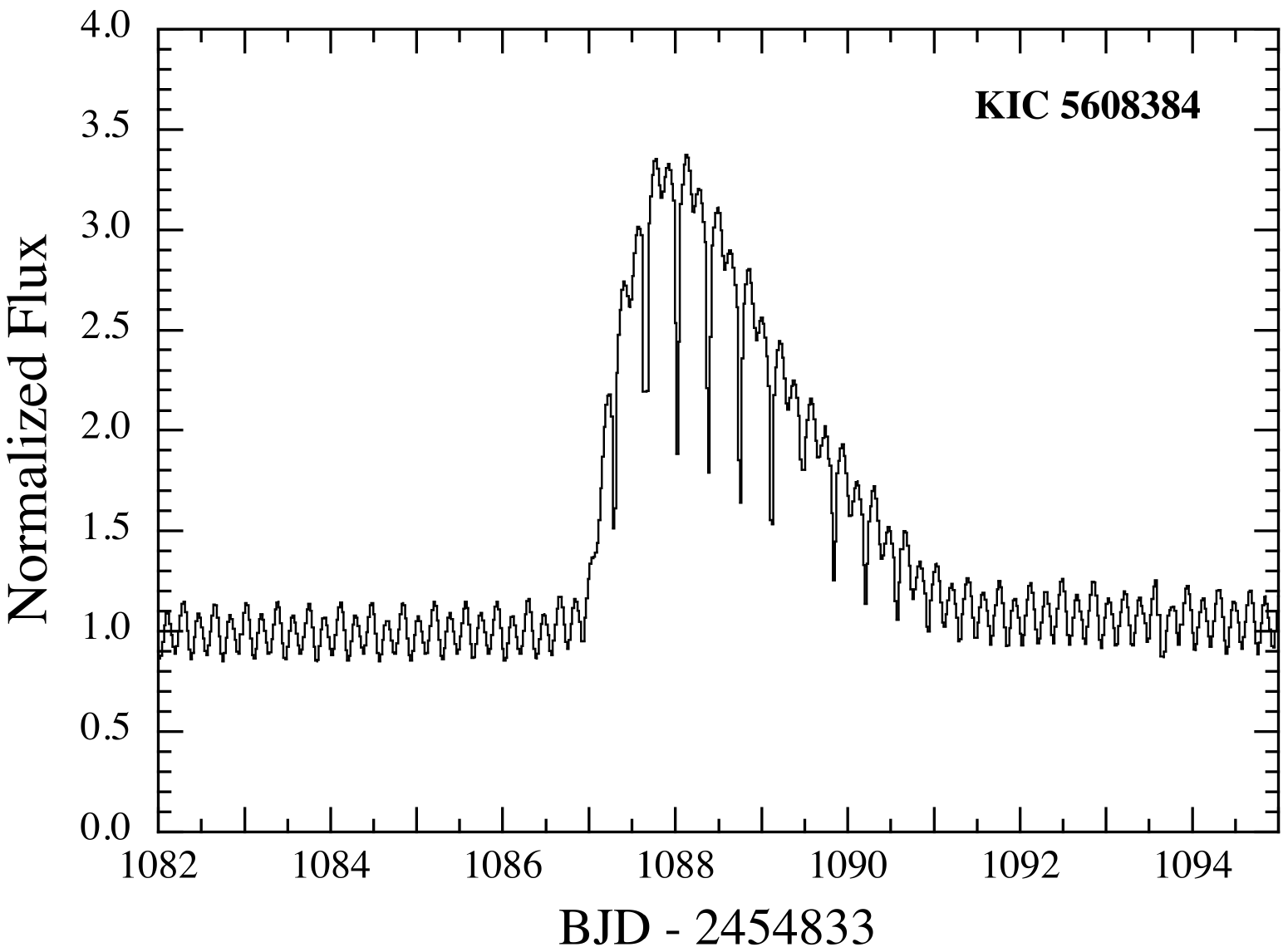}
\caption{{\em Kepler} lightcurve of KIC 5608384. The top panel shows the entire 1500-day lightcurve from all four years of {\em Kepler} observations. Only a single large flaring event is seen near day 1100.  The bottom panel is a zoom-in of 13 days around the time of the outburst. Now the characteristic shape of a CV outburst is evident as are the eclipses of at least a portion of the accretion disk.  Also visible in the bottom panel are the 8.7-hour ellipsoidal light variations with a $\sim$12\% amplitude.  }
\label{fig:lightcurve} 
\end{center}
\end{figure}

\noindent
The {\em Kepler} mission, with its exquisite photometric precision (\citealt{borucki10}; \citealt{batalha11}) has revolutionized stellar astronomy.  During the main {\em Kepler} mission, the fluxes from some 150,000 stars were monitored with a typical cadence of 1/2 hr nearly continuously for four years.  The instrument has a broad spectral response extending from 0.43 $\mu$m to 0.89 $\mu$m.  Thousands of exoplanets (e.g., \citealt{borucki10}; \citealt{batalha13}) as well as eclipsing binaries (\citealt{prsa11}; \citealt{slawson11}; \citealt{matijevic12}) have been discovered with these data.  

\begin{figure}
\begin{center}
\includegraphics[width=0.99\columnwidth]{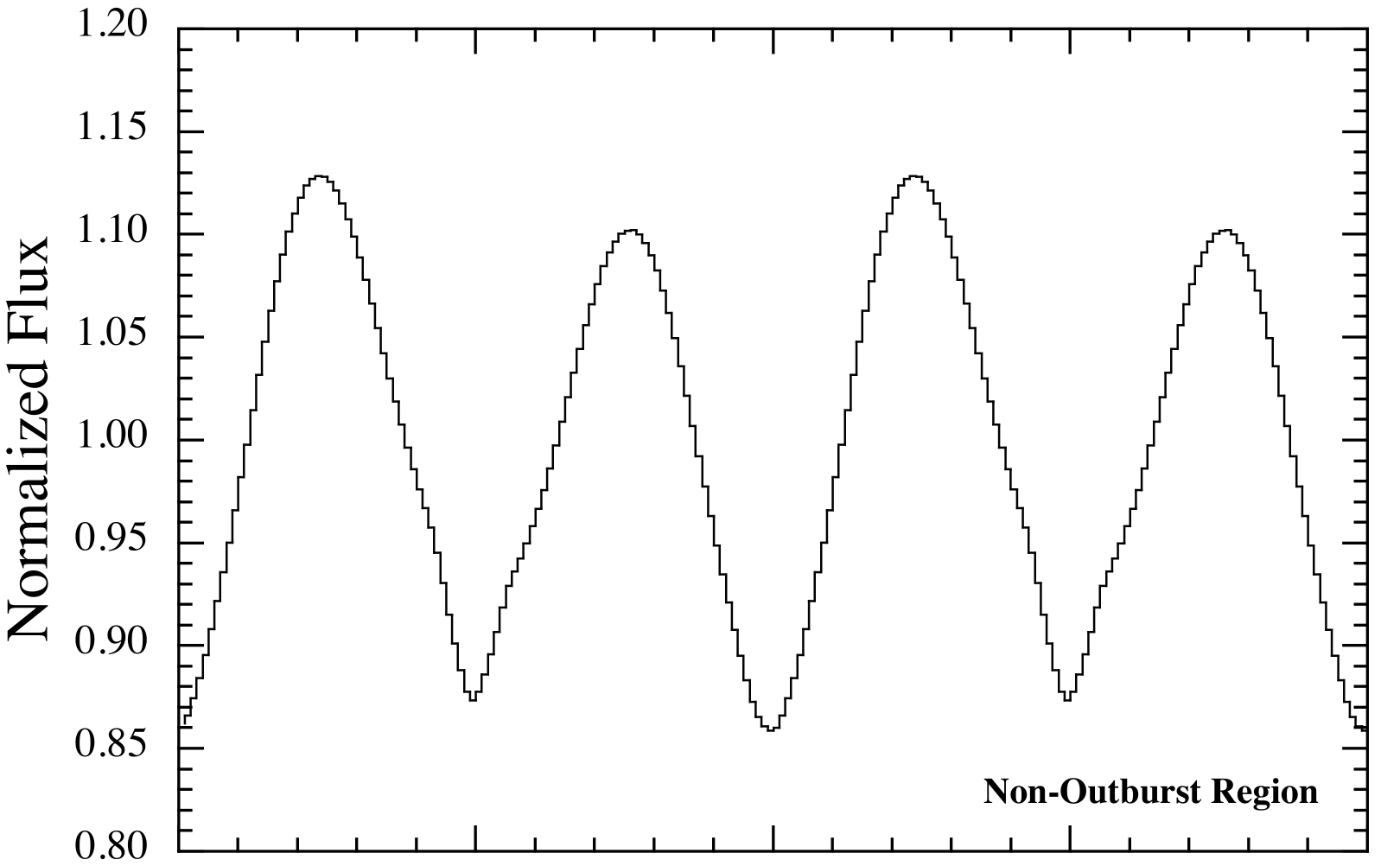} \hglue0.112cm
\includegraphics[width=0.977\columnwidth]{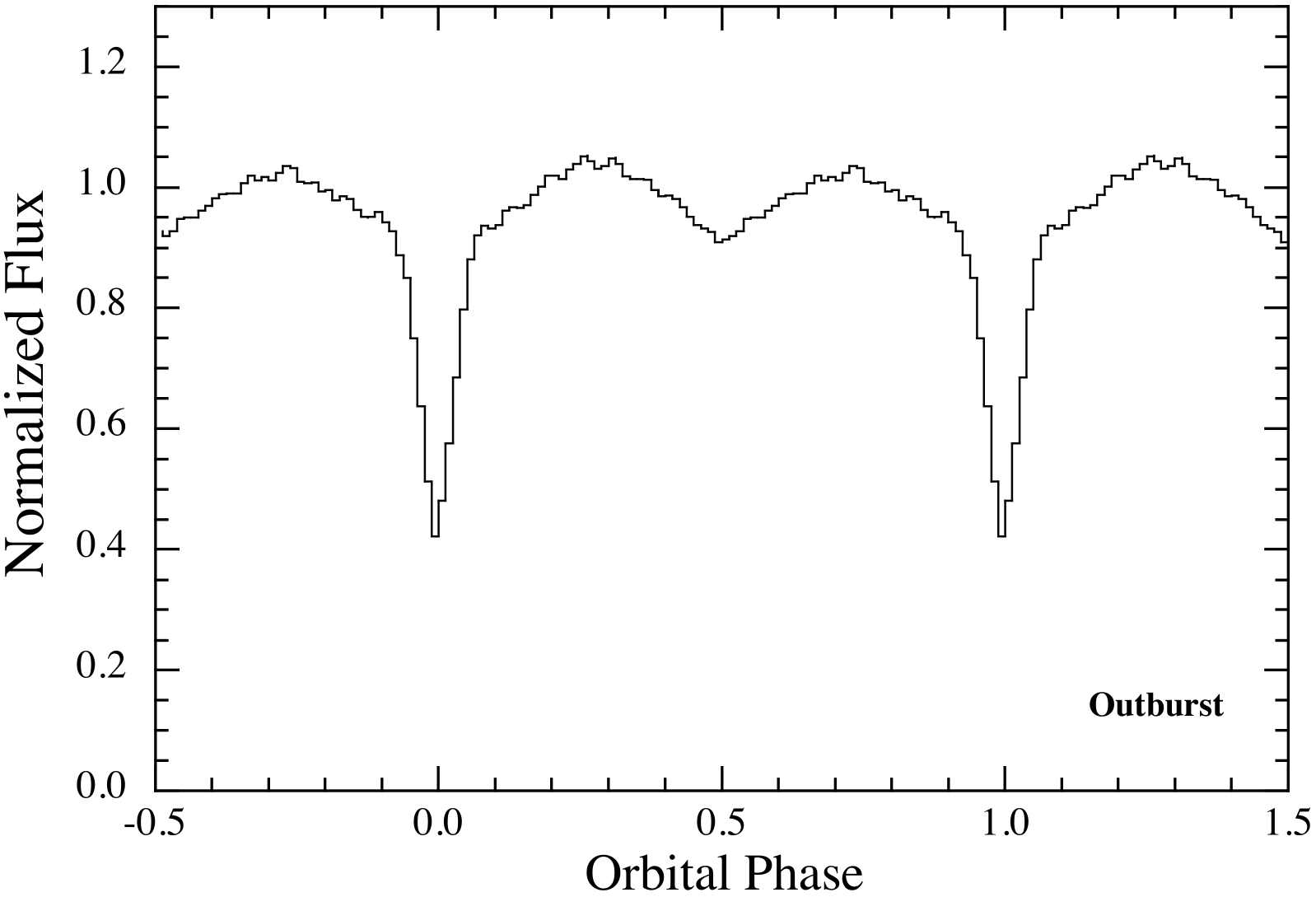} 
\caption{Folded lightcurve for KIC 5608384 based on the {\em Kepler} data set. The data are folded about the 8.7-hour orbital period, and then repeated for a second orbital cycle to better show the pattern. The top panel indicates how the lightcurve is dominated by ellipsoidal light variations with a $\sim$12\% amplitude and two maxima and two minima per orbit. Here, all the {\em Kepler} data are used except for the 5 days around the time of outburst.  The bottom panel shows the folded lightcurve during the outburst after normalizing the data to the upper profile of the outburst.  The partial eclipse of the accretion disk during the outburst can be seen with a depth of $\gtrsim$50\%.  Orbital phase zero is defined at the inferior conjunction of the K star. Note the large difference in the vertical scales between the two plots.}
\label{fig:fold} 
\end{center}
\end{figure}

As part of an ongoing effort to identify unusual objects in the {\em Kepler} main field, one of us (T.\,J.) visually inspected all $\sim$200,000 of the {\em Kepler} lightcurves (see, e.g., \citealt{rappaport18}).  One unusual object that was found was KIC 5608384, a red star previously cataloged as a variable star (V0754 Lyr, \citealt{kryachko10}; ASASSN-V J191223.18+404952.8, \citealt{shappee14}, \citealt{kochanek17}), which exhibits one CV-like outburst during the four-years of observation. The full four-year {\em Kepler} lightcurve is shown in Fig.~\ref{fig:lightcurve}.  Note the large increase in flux near day BKJD = 1088 (defined as BJD - 2454833). This region of the outburst is shown in the bottom panel of Fig.~\ref{fig:lightcurve}. Large periodic dips in flux with a period of 8.7 hours can clearly be seen during the outburst. However, the same period can also be seen during the full four years of the {\em Kepler} mission, as a sinusoidal modulation of much lower amplitude.  The latter is taken to be the ellipsoidal light variations (at twice the orbital frequency) of the binary system. 

The basic photometric and kinematic properties of this object are summarized in Table \ref{tbl:mags}.  We make use of various items in this table throughout the paper.

\begin{table}
\centering
\caption{Photometric Properties of KIC 5608384}
\begin{tabular}{lc}
\hline
\hline
Parameter &
KIC 5608384 \\
\hline
RA (J2000) & 19:12:23.30   \\  
Dec (J2000) &  $40$:49:53.44  \\  
$K_p$ & 14.922  \\
$G$$^a$ & $14.881 \pm 0.006$ \\
$G_{\rm BP}$$^a$ & $15.592\pm 0.022$ \\
$G_{\rm RP}$$^a$ & $14.064 \pm 0.018$ \\
J$^b$& $12.941 \pm 0.021$ \\
H$^b$ & $12.313 \pm 0.021$ \\
K$^b$ & $12.137 \pm 0.020$ \\
W1$^c$ & $12.070 \pm 0.023$ \\
W2$^c$ & $12.073 \pm 0.024$ \\
W3$^c$ & $11.790 \pm 0.157$ \\
W4$^c$ & $> 9.5$ \\
$T_{\rm eff}$$^a$ (K) & $4511 \pm 260$ \\
Distance (pc)$^a$ & $366.8 \pm 2.7$  \\   
$\mu_\alpha$ (mas ~${\rm yr}^{-1}$)$^a$ & $11.62 \pm 0.04$  \\ 
$\mu_\delta$ (mas ~${\rm yr}^{-1}$)$^a$ &  $14.45 \pm 0.04$  \\ 
\hline
\label{tbl:mags}  
\end{tabular}

{\bf Notes.} (a)  Gaia DR2 \citep{lindegren18}. (b) 2MASS archive \citep{skrutskie06}. (c) WISE archive \citep{cutri13}. 
\end{table}

To visualize and determine the photometric properties of KIC 5608384 more precisely, we folded all the {\em Kepler} data during quiescence (see Fig.~\ref{fig:lightcurve}; i.e., except for the several days surrounding the outburst).  The results are shown in the top panel of Fig.~\ref{fig:fold}. Here, the 8.7 hour ellipsoidal light variations are quite apparent with their two characteristic maxima per orbital cycle. Note that orbital phase zero is defined as the inferior conjunction of the K star (i.e., superior conjunction of the white dwarf). The bottom panel displays the folded lightcurve during the outburst.  To produce this plot, we fitted a function to the upper profile of the 4-day long flare (bottom panel of Fig.~\ref{fig:lightcurve}), and use this function to renormalize the data to unity.  The resulting plot shows the partial eclipses of the accretion disk with a depth of $\gtrsim$50\% (including the steady contribution from the K star). The observed eclipse duration is $\pm 0.06$ orbital phases, which corresponds to 1.05 hours. Therefore the width has been affected (i.e., increased) by the 1/2 hour cadence of the {\rm Kepler} observation.

An examination of the folded {\em Kepler} lightcurve (see Fig.~\ref{fig:fold}) reveals that the ellipsoidal light variations are not the only contribution to the flux modulations. We arrive at this conclusion based on the fact that the peak heights of two consecutive maxima are different. This effect is likely produced by spots on the K star which, in turn, are likely to be synchronously corotating with the orbit. To more quantitatively study the behavior of the spot(s), we track the phases of the sinusoids varying at $\omega t$ and $2 \omega t$ \citep{balaji15}.  We therefore fit a function of the following form to the {\em Kepler} data in five-day  segments: 
\begin{equation}
f(t)=A+B\sin(\omega t)+C\cos(\omega t)+D\sin(2\omega t)+E\cos(2\omega t) ~,
\end{equation}
where $A$, $B$, $C$, $D$, and $E$ are five parameters to be fit and $\omega$ = $2\pi/P_{\rm orb}$. The $2\omega$ terms found from the $D$ and $E$ coefficients largely track the ellipsoidal light variations, while the $\omega$ terms found from $B$ and $C$ are mostly dominated by the presence of one or more starspot(s).  After the fit is done, the fitting window is shifted by one day in the data train, to produce a type of running average of the fitted parameters. 

\begin{figure}
\begin{center}
\includegraphics[width=0.98 \columnwidth]{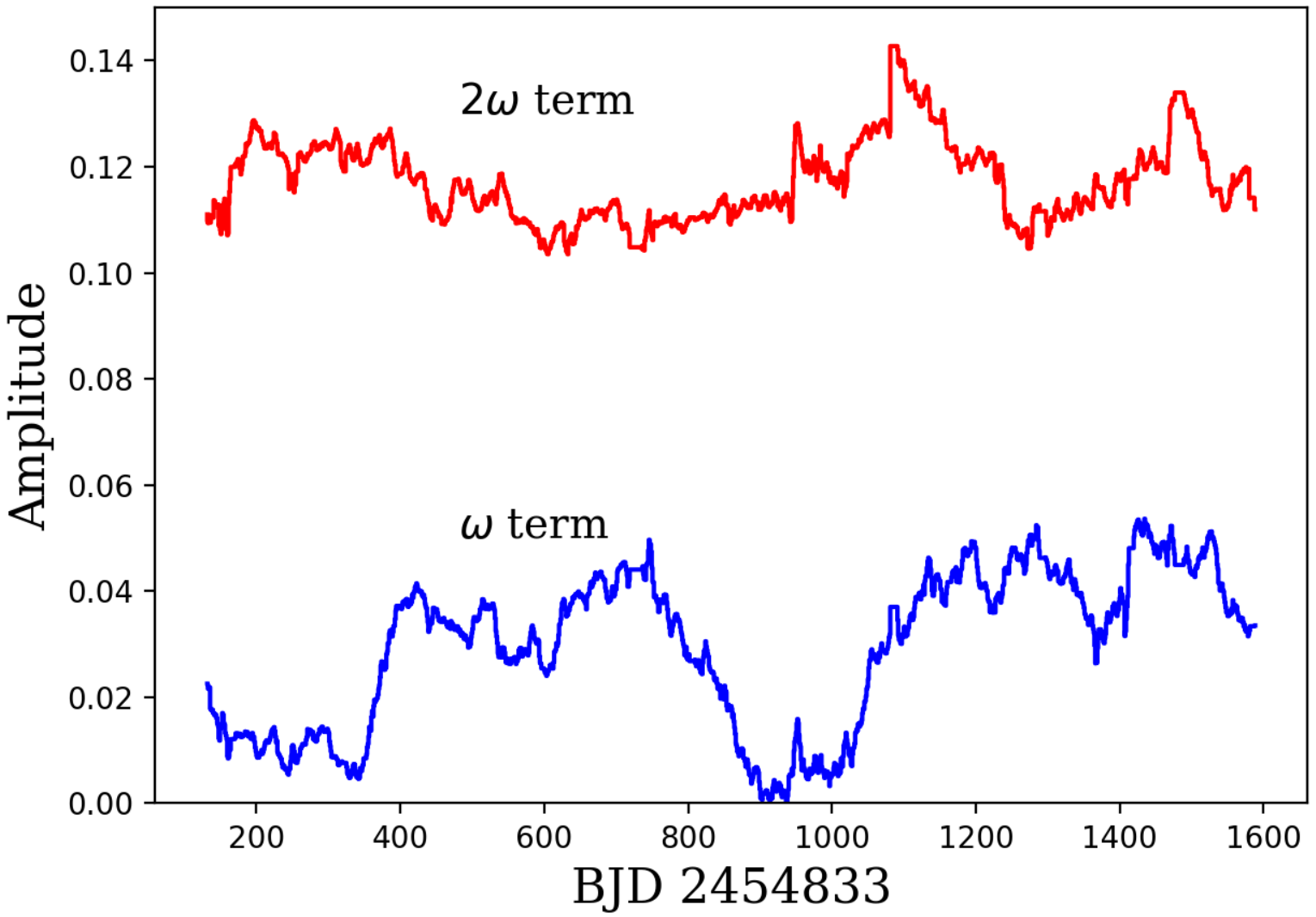} \hglue-0.08cm
\includegraphics[width=0.997 \columnwidth]{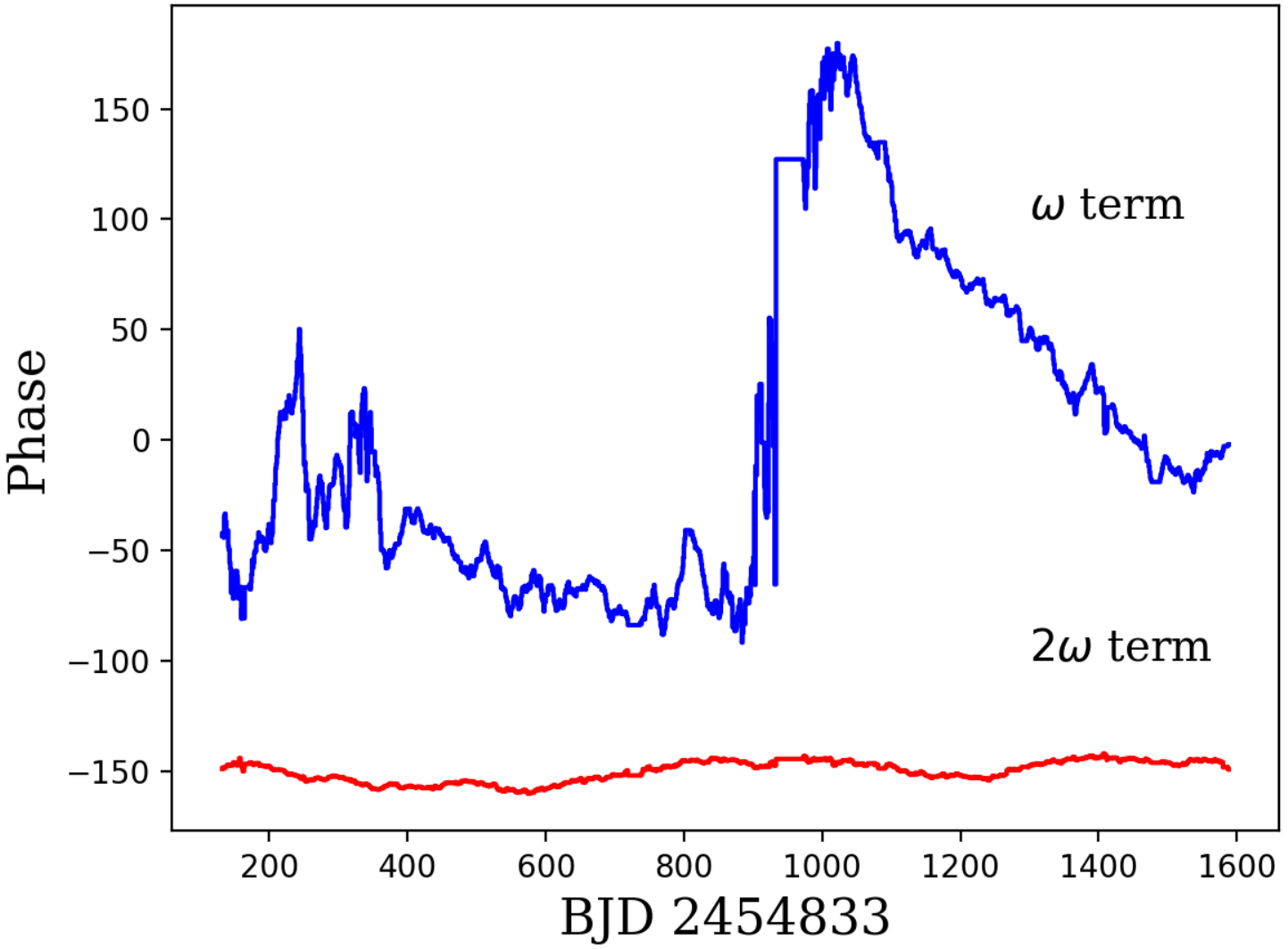} 
\caption{Phase tracking the 8.7-hour modulations of KIC 5608384 (see Sect.~\ref{sec:Kepler} for details). Top panel: amplitude of the brightness variations at $\omega$ and at $2\omega$ ($\omega$ is the orbital frequency). The $2\omega$ term represents primarily the ellipsoidal light variations, while the $\omega$ term is sensitive to spots on the K star that are nearly corotating in synchronism with the orbit.  Bottom panel: phase of the brightness variations at $\omega$ and at $2\omega$.  Note that both the phase and amplitude of the $2\omega$ term, representing the ELVs, are relatively constant for four years while the phase and amplitude of the $\omega$ term, representing spots on the K star, are highly variable.}
\label{fig:phase_tracking} 
\end{center}
\end{figure}

The top panel in Fig.~\ref{fig:phase_tracking} shows the fitted amplitudes of the $\omega$ term and $2\omega$ term vs.~time.  Note that the amplitude of the $2 \omega$ term is nearly constant at $\simeq 11.8\%$, as expected if this term is dominated by the immutable ellipsoidal light variations (`ELVs').  By contrast, the amplitude of the $\omega$ term varies between 0\% and 4\%, which is presumably caused by the evolution of the spot(s) on the K star with timescales of $\sim$100 days.  The bottom panel displays the evolution of phases ($\phi_1 = \tan^{-1}[B/C]$ and $\phi_2 = \tan^{-1}[D/E]$) of the two terms with time. As expected, the $2\omega$ term remains nearly constant in phase, varying by only a few degrees.  By contrast, the phase of the $\omega$ term varies dramatically with time, being relatively constant for the first 700 days of the {\em Kepler} mission, then undergoing a large phase shift of $\sim$$250^\circ$, and finally following a downward trend in the phase plot.  Both the amplitude and phase behavior of the $\omega$ term are suggestive of a spot or spots that are evolving in time (see discussion in \citealt{balaji15}).

\begin{figure}
\begin{center}
\includegraphics[width=0.99 \columnwidth]{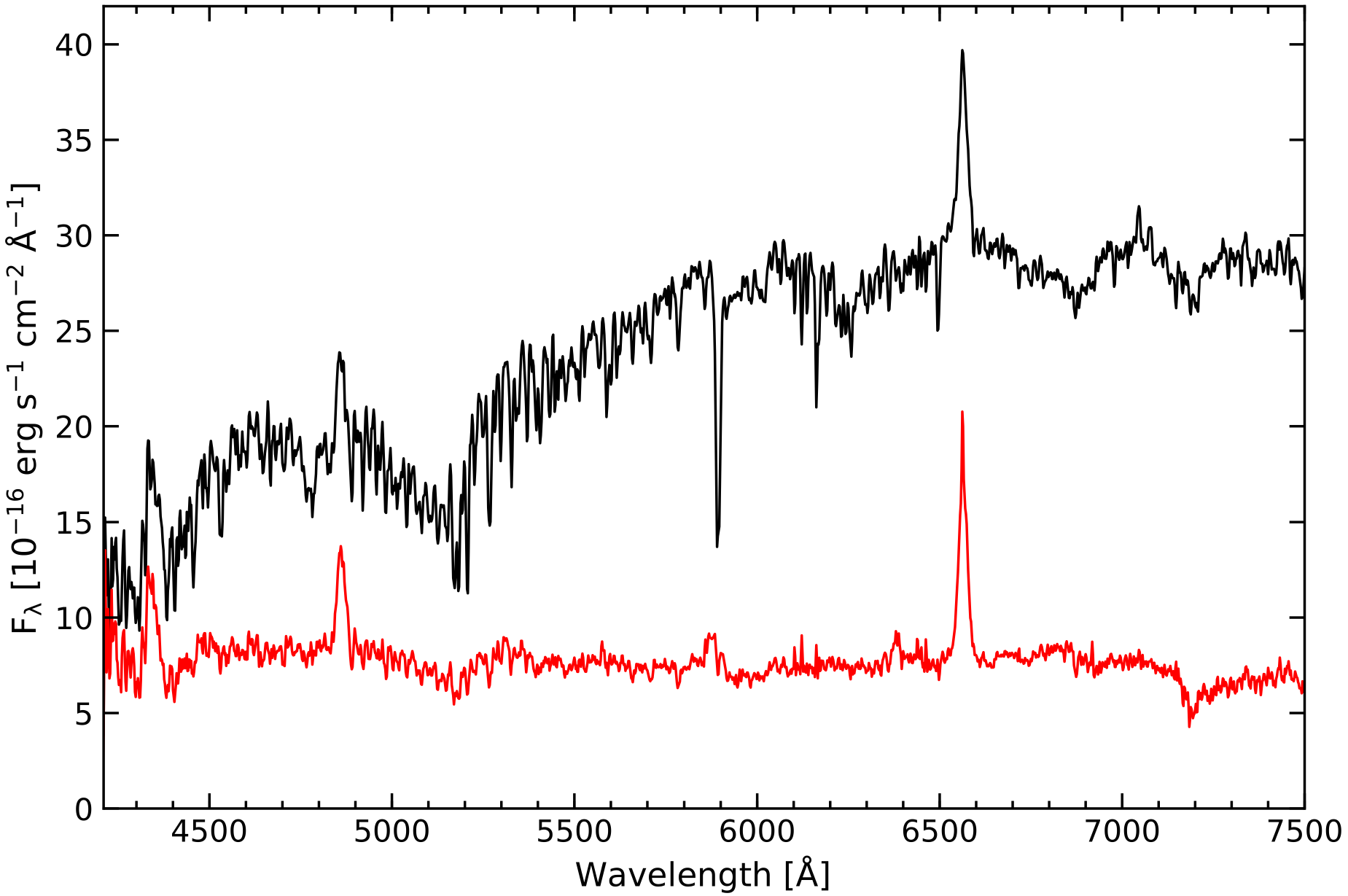}  
\includegraphics[width=0.99 \columnwidth]{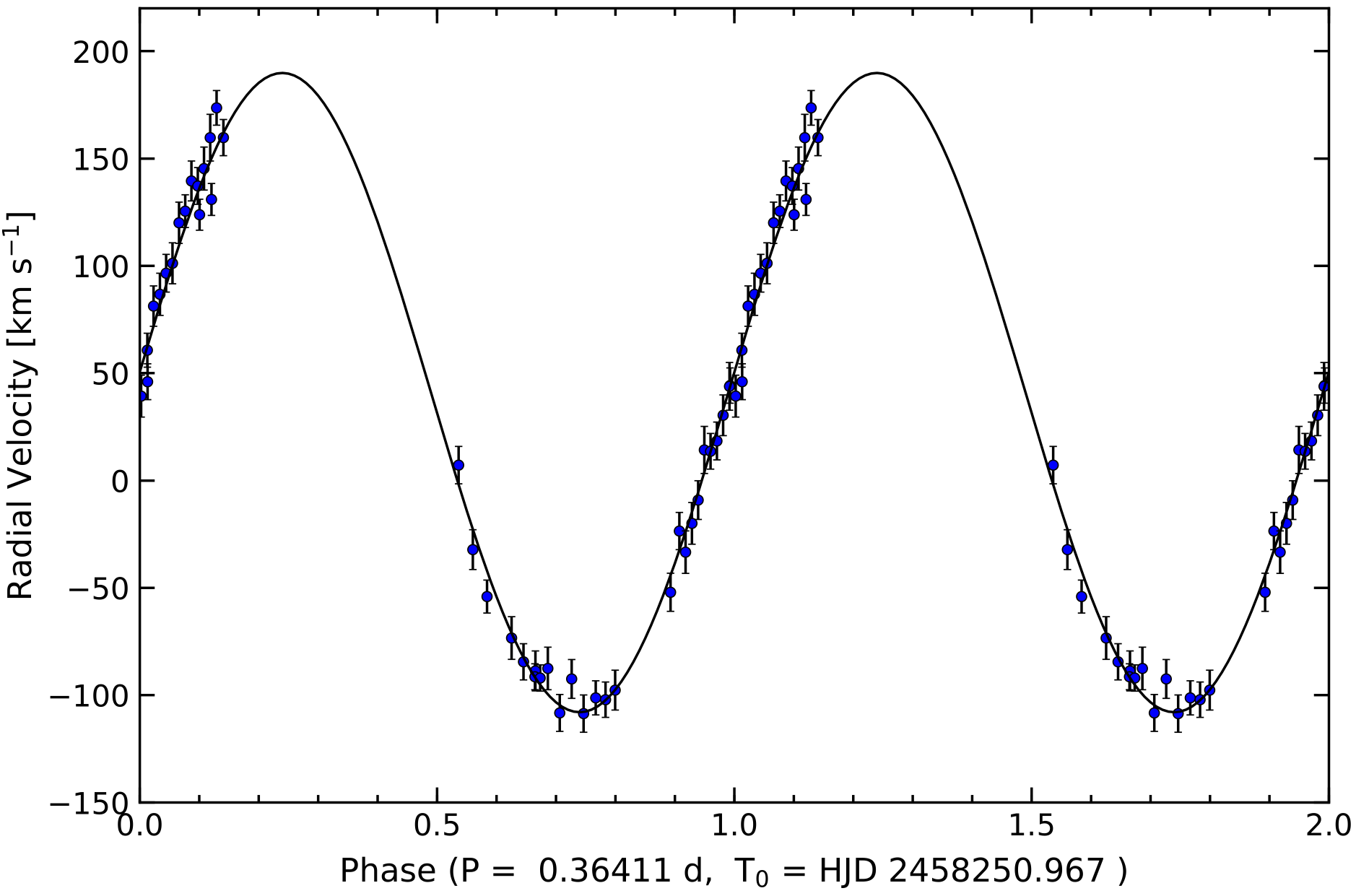} 
\includegraphics[width=0.99 \columnwidth]{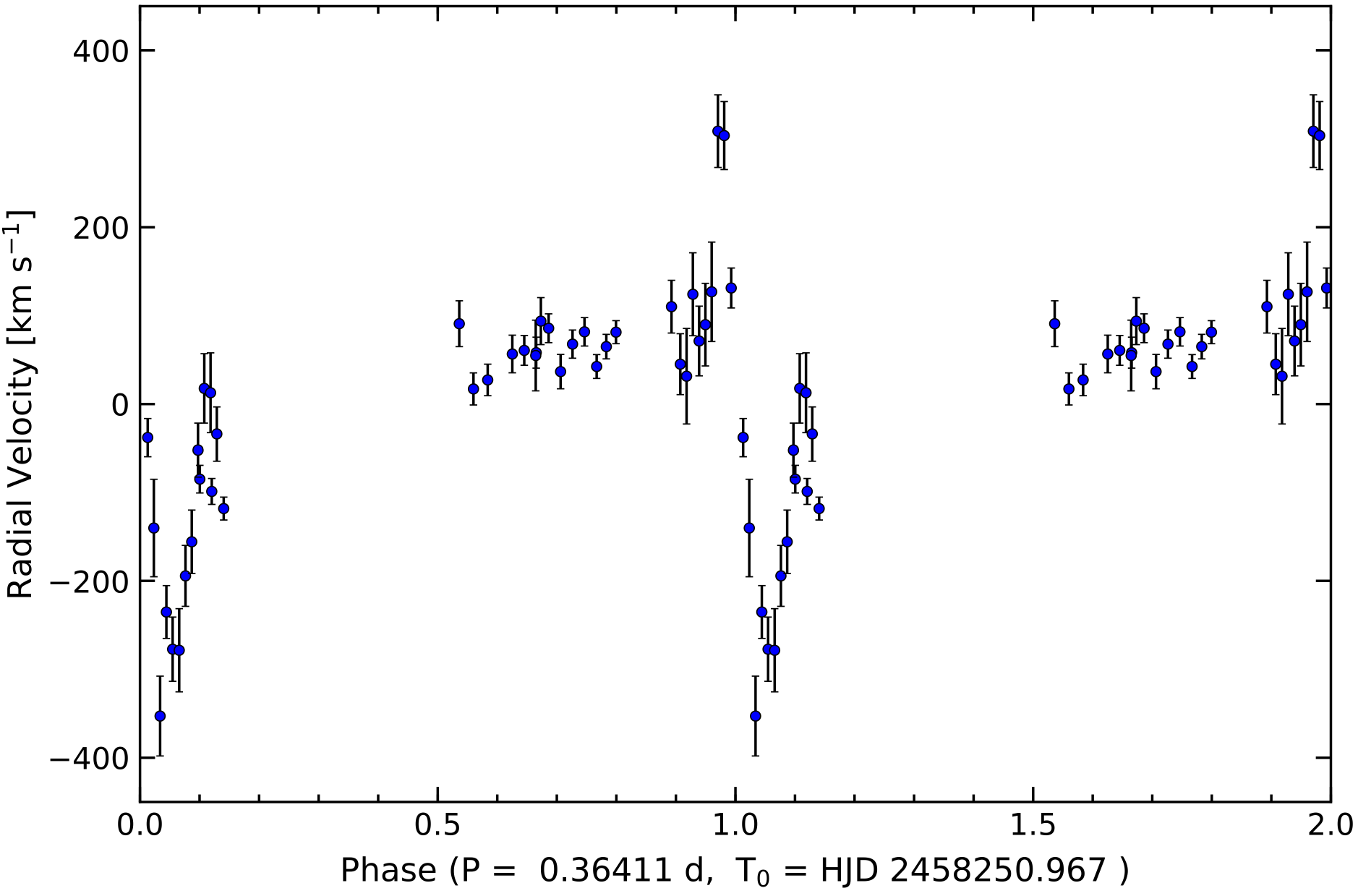} 
\caption{Ground-based radial velocity data (see Sect.~\ref{sec:spectra}).  Top panel: mean spectrum of KIC 5608384, shifted to the rest frame of the
secondary star before averaging.  The lower red curve results from subtracting the spectrum of Gliese 638.  Middle panel: absorption-line velocities, folded on the orbital period with the best-fitting sinusoid superposed.  All data are repeated for an extra cycle for continuity.  Bottom panel: emission-line velocities folded on the orbital period.  Note the strong rotational disturbance around phase zero, where the disk is being eclipsed.} 
\label{fig:spectrum} 
\end{center}
\end{figure}

\begin{figure*}
\begin{center}
\includegraphics[width=0.99\textwidth]{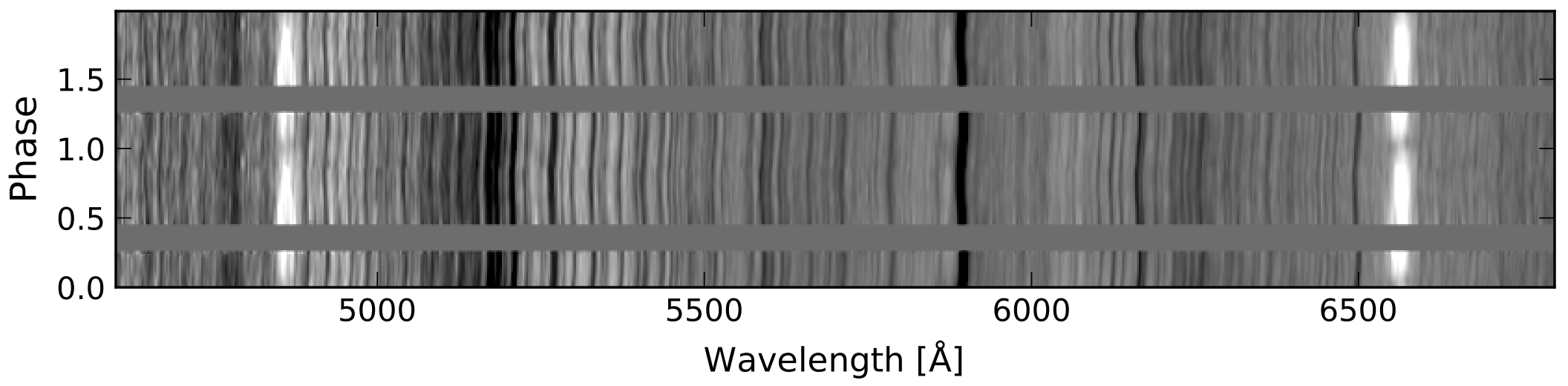} \vglue0.4cm
\includegraphics[width=0.6\textwidth]{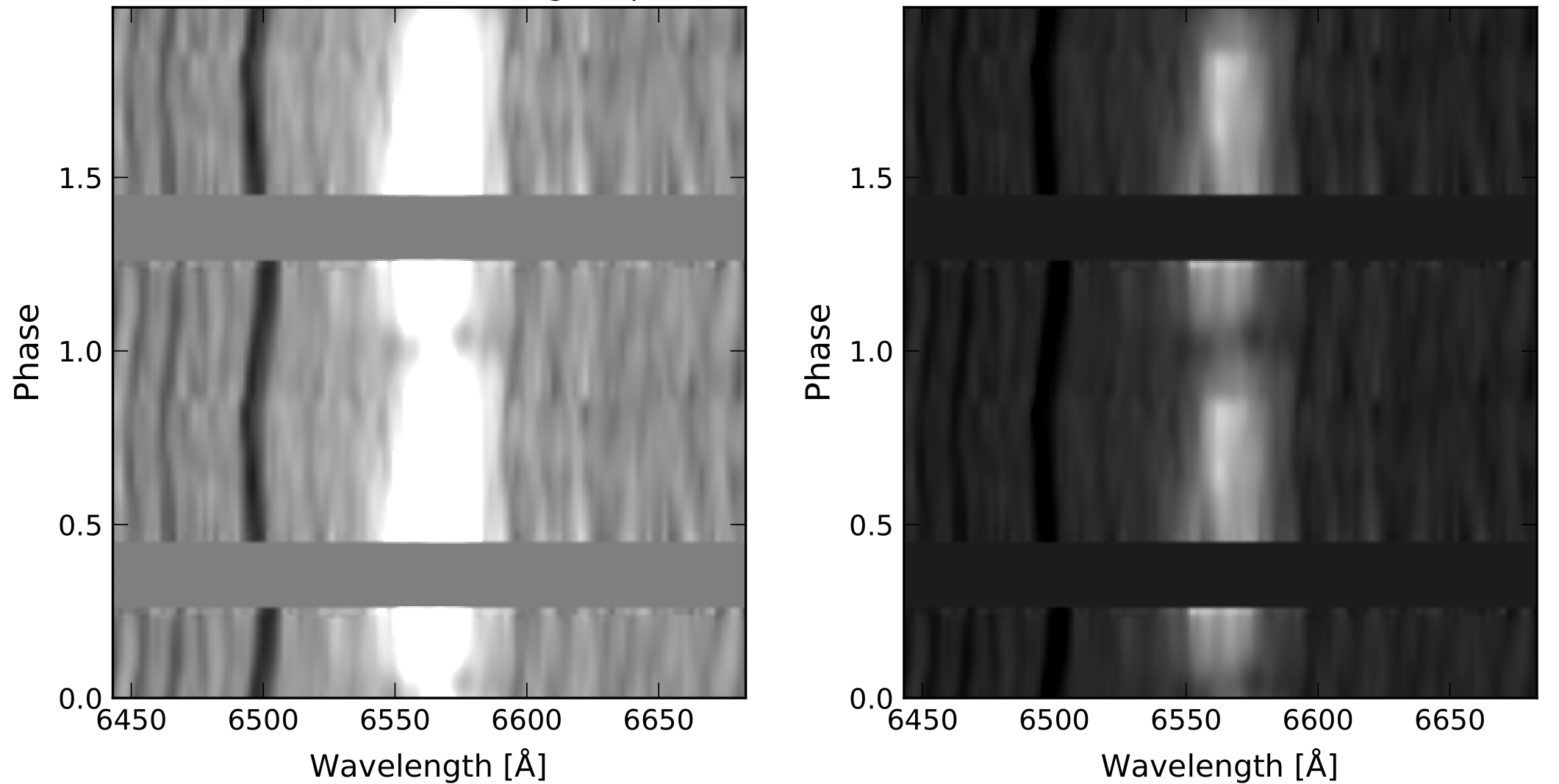} 
\caption{Top panel: trailed spectrum of KIC 5608384 as a function of orbital phase, with a second cycle shown for continuity. The grey scale is set so that lighter and darker regions correspond to emission and absorption lines, respectively.  Bottom panel: trailed spectra zoomed in around the H$\alpha$ line with two different contrasts to increase the discernible dynamic range.  Note that there appears to be H$\alpha$ absorption from the underlying stellar photosphere that complicates the profile of the relatively weak emission line.}
\label{fig:trailed} 
\end{center}
\end{figure*}

\section{Ground-Based Spectra} 
\label{sec:spectra}

We obtained 41 spectra of KIC 5608384 with the McGraw-Hill 1.3-m telescope at MDM Observatory, on Kitt Peak, Arizona, operated remotely, during four nights in 2018 May, and also two exposures with the 2.4\,m Hiltner telescope in 2018 June. A $2048 \times 2048$ SITe CCD mounted on the modular 
spectrograph\footnote{http://mdm.kpno.noao.edu/index/Instrumentation.html} (which can be used with either telescope) gave a useful range from 4300 to 7500 \AA, vignetted toward the ends, a dispersion of 2 \AA\ pixel$^{-1}$, and a slightly undersampled resolution of  3.5 \AA\ FWHM.  The observing and data-reduction procedures were basically as described by \citet{julesxray3}.  We used the [OI] $\lambda$5577 night-sky line found in every spectrum to shift the zero point of the wavelength calibration, and applied a flux calibration derived from standard stars observed during twilight. The same physical slit width was used on both telescopes, the projected widths being $1.04''$ and $1.92''$ on the 2.4\,m and 1.3\,m respectively. Uncalibrated light losses at the rather narrow slit made the flux calibration uncertain by $\sim$0.3 mag, as estimated from the scatter of the comparison stars.

\begin{table}
\centering
\caption{Radial Velocities for KIC 5608384$^a$}
\begin{tabular}{lcc}
\hline
\hline
Time & RV (absorption) & RV (emission)    \\
(MJD$_{\rm tdb}$) & (km s$^{-1}$) & (km s$^{-1}$) \\
\hline
58249.7901 &  $-101  \pm 8$ & $+42 \pm13$ \\
58249.7961  & $-102 \pm 8$  & $+65 \pm 14$ \\
58249.8020  & $-98 \pm 9$ & $+81 \pm 13$ \\ 
58249.9116  & $+124 \pm 7$ & $-85 \pm 16$ \\
58249.9189  & $+131 \pm 7$ & $+99 \pm 15$ \\
58249.9262  & $+160 \pm 8$ & $-118 \pm 13$ \\  
58250.8309  & $-73 \pm 10$ & $+56 \pm 21$ \\   
58250.8382  & $-84 \pm 8$ & $+61 \pm 17$ \\   
58250.8455  & $-89 \pm 9$ & $+58 \pm 18$ \\
58250.8531  & $-88 \pm 10$ & $+86 \pm 16$ \\   
58250.8604  & $-108 \pm 9$ & $+37 \pm 19$ \\   
58250.8677  & $-92 \pm 9$ & $+68 \pm 16$ \\  
58250.8750  & $-109 \pm 9$ & $+82 \pm 16$ \\   
58250.9647  & $+44 \pm 8$ & $+131 \pm23$ \\  
58250.9721  & $+46 \pm 8$ & $-38 \pm 21$ \\   
58251.8908  & $+7 \pm 9$ & $+91 \pm 26$ \\  
58251.8995  & $-32 \pm 9$ & $+17\pm 18$ \\    
58251.9082  & $-54 \pm 8$ & $+27 \pm18$ \\  
58252.7488  & $-52 \pm 9$ & $+110 \pm 30$ \\  
58252.7542  & $-24 \pm 9$ & $+45 \pm 34$ \\   
58252.7580  & $-33 \pm 10$ & $+31 \pm 54$ \\   
58252.7619  & $-20 \pm 10$ & $+124 \pm 47$ \\   
58252.7657  & $-9 \pm 9$ & $+71 \pm 39$ \\   
58252.7695  & $+14 \pm 11$ & $+90 \pm 47$ \\   
58252.7733  & $+14 \pm 8$ & $+127 \pm 56$ \\   
58252.7772  & $+18 \pm 9$ & $+309 \pm 41$ \\   
58252.7810  & $+30  \pm 9$ & $+304 \pm 38$ \\   
58252.7849  & $+43.9 \pm 11.1$ & ... \\   
58252.7887  & $+39.3 \pm 9.8$ & .... \\  
58252.7925  & $+60.7  \pm 7.9$ & ... \\  
58252.7963  & $+81  \pm 9$ & $-140 \pm 55$ \\   
58252.8002  & $+87  \pm 10$ & $-353 \pm 45$ \\  
58252.8040  & $+97  \pm 9$ & $-235 \pm 30$ \\  
58252.8079 & $+101 \pm 	0$ & $-277 \pm 36$ \\   
58252.8119 & $+120 \pm 10$ & $-278 \pm 47$ \\   
58252.8157 & $+125 \pm 8$ & $-194 \pm 34$ \\  
58252.8195 & $+140 \pm 9$ & $-156 \pm 36$ \\
58252.8234 & $+137 \pm 9$ & $-52 \pm 30$ \\   
58252.8272 & $+145 \pm 10$ & $+18 \pm39$ \\  
58252.8310 & $+160 \pm 11$ & $+13 \pm 45$ \\   
58252.8349  & $+174 \pm 8$ & $-34 \pm 31$ \\  
\hline
\hline
Time ($\phi =0$) & $K$ & $\gamma$  \\
(BJD$_{\rm tdb}$)  & (km s$^{-1}$) & (km s$^{-1}$)  \\
\hline
58250.5994 & $148.6 \pm 2.8$ & $40.3 \pm 2.1$   \\
\hline 
\label{tbl:RVs}  
\end{tabular}

{\bf Notes.} (a) Radial velocities of the late-type absorption spectrum and H$\alpha$ line (when measurable).
The time given is the barycentric Julian date of mid-integration, minus 2,440,000.0, on the UTC system.
\end{table}

We measured absorption-line velocities using the IRAF task {\tt xcsao} \citep{kurtzmink}, which implements the \citet{tonrydavis} cross-correlation algorithm.  For a known-velocity template spectrum, we used a sum of 76 spectra of K-type velocity standards taken with the same instrument;
the individual spectra in the sum were shifted to their rest frames before summation. We cross-correlated the spectral range from 5000 to 6500 \AA , but ignored a window from 5850 to 5950 \AA, to avoid contamination of the photospheric absorption by both the He I $\lambda$5876 emission line, and any interstellar contribution to NaD absorption.  Note that the cross-correlation method does not use any individual line, but considers the entire spectral range simultaneously.  

H$\alpha$ was the strongest emission line, and was the only one strong enough to give useful velocities. To measure emission velocities we used an 
algorithm developed by \citet{sy80}, which convolves the line profile with an odd-parity function, and takes the point at which the convolution is zero as the line center.  

Figure \ref{fig:spectrum} (top panel) shows the mean spectrum, shifted into the rest frame of the secondary star before averaging (see below).  The bulk 
of the light comes from a late-type star, which we classify as K7 $\pm$ 2 subclasses.  The lower (red) trace shows the spectrum after subtraction
of a spectrum of Gliese 638 (classified K7.5 by \citealt{keenan89}), obtained with the same instrument and scaled to approximately match the features in the spectrum.  The match is not perfect, as spectral-type standards evidently differ even at the same subtype, but the residuals clearly show the emission lines.

Table \ref{tbl:RVs} gives radial velocities of the emission and absorption lines.  The absorption-line velocities are strongly modulated on the 0.364111-d photometric period (Fig.~\ref{fig:spectrum}; middle panel) proving that this period is orbital.  There is no evident eccentricity.  Fitting 
a sinusoid of the form $$v(t) = \gamma + K \sin[2 \pi (t - T_0) / P],$$ to the absorption velocities via least-squares, and leaving $P$ fixed
at 0.364111 d, yields 
\begin{eqnarray*} 
\gamma &=& 41 \pm 3\ {\rm km\ s^{-1}},\\ 
K &=& 149 \pm 4\,{\rm km\ s^{-1}},\\
{\rm and\ } T_0 &=& {\rm BJD\ } 2458252.7842 \pm 0.0014,\\ 
\end{eqnarray*}
where $T_0$ is the time of the eclipse of the white dwarf (or the ascending node of the RV curve) in the UTC time system. The RMS residual of the fit is 9 km s$^{-1}$. These results are summarized at the bottom of Table \ref{tbl:RVs}.

The velocities of H$\alpha$ (Fig.~\ref{fig:spectrum}; bottom panel) are much less orderly, but show a pronounced Rossiter-McLaughlin effect
\citep{rossiter24, mclaughlin24} -- rapid excursions to the red and then to the blue as an eclipse covers first the blueshifted and then the redshifted sides of a rapidly rotating accretion disk.

In Fig.~\ref{fig:trailed} we show two orbital phase-interpolated representations of the spectra. The top panel shows nearly the full wavelength range and the grey-scale is set to highlight absorption vs.~emission lines.  The bottom panel zooms in around the H$\alpha$ line and shows two different contrast levels to enhance the discernible dynamic range. These plots were produced by setting up a grid of 100 evenly-spaced phases around the orbit, computing a spectrum at each phase by averaging the spectra taken near that phase with a Gaussian weighting function in phase, and finally stacking these together to make a two-dimensional greyscale image.  The spectra were rectified (divided by a low-order polynomial fit to normalize the continuum) prior to averaging.  The featureless horizontal bands are phases for which we have no data.  It is easy to see the movement of the absorption lines, and the near-disappearance of the H$\alpha$ emission around phase zero is also very clear.  The Rossiter-McLaughlin disturbance is less obvious in this representation, and the orbital modulation of the H$\alpha$ velocity is also rather obscure, in part because of the gaps in phase coverage.

\section{Swift and GALEX ultraviolet photometry}
\label{sec:UV}

KIC~5608384 has been detected by the deep \textit{GALEX} near-ultraviolet survey of the \textit{Kepler} field at $NUV=18.588\pm0.017$ \citep{olmedoetal2015}, indicating an ultraviolet excess over the flux from the donor star. To further characterise the ultraviolet flux of KIC~5608384, we obtained \textit{Swift} ToO observations in the broadband UVOT uvw2, uvm2, and uvw1 bands (Table\,\ref{tbl:uvmags}) in November 2018. The \textit{Swift} observation detects KIC~5608384 at a near-ultraviolet flux level comparable to the \textit{GALEX} data (Table\,\ref{tbl:uvmags}). 

\begin{figure}
\begin{center}
\includegraphics[width=1.01 \columnwidth]{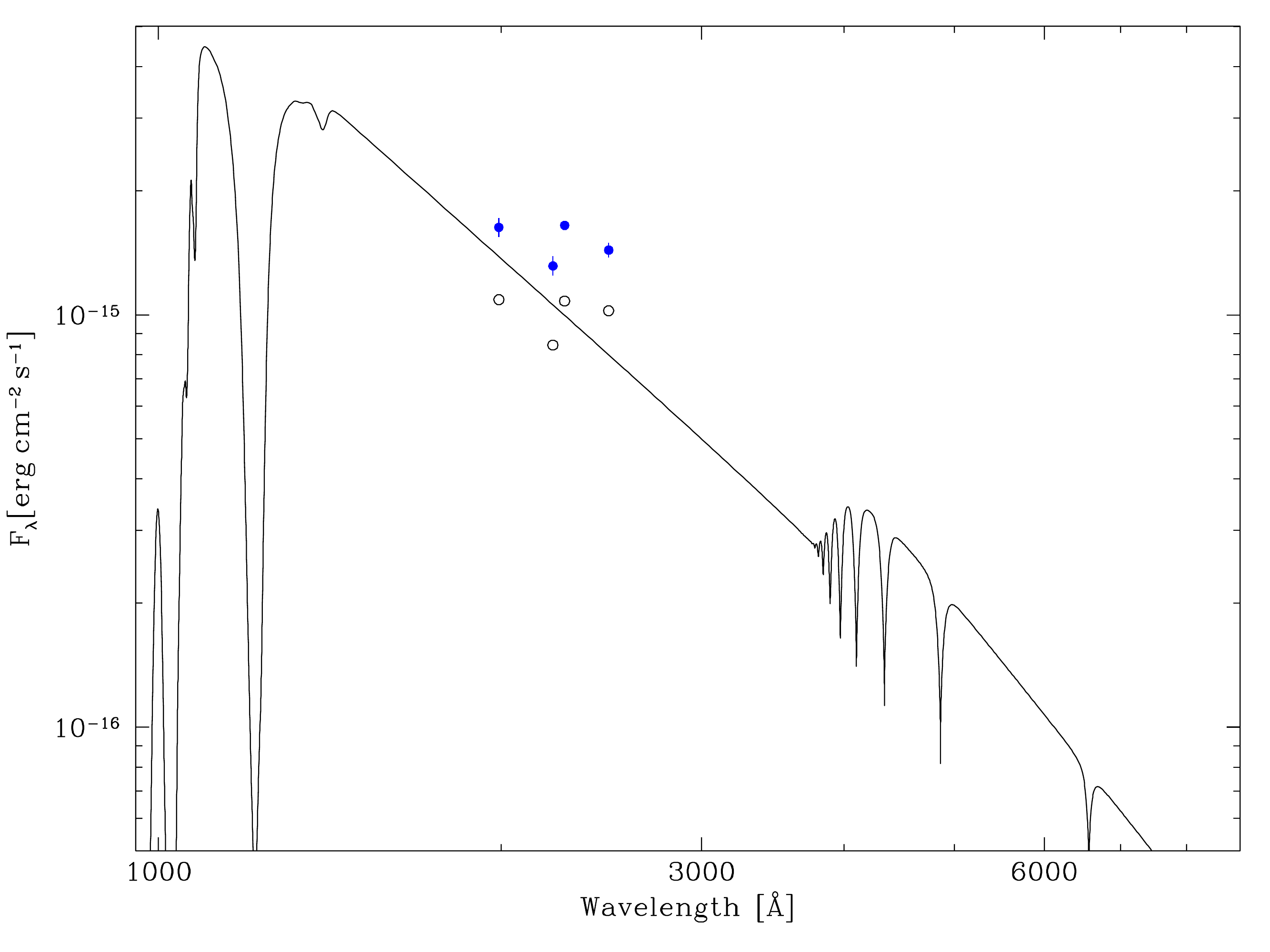} 
\caption{Ultraviolet fluxes from KIC~5608384 from the \textit{GALEX} and \textit{Swift} photometry (Table \ref{tbl:uvmags}).  The observed fluxes are shown in open circles (black), and dereddened with $E(B-V)=0.05$ in filled (blue) circles.  Adopting a white dwarf mass of $0.47\,M_\odot$ (see Sect.~\ref{sec:mcmc}), the distance of 367\,pc (Table \ref{tbl:mags}), a mass-radius relation for the white dwarf, and that the white dwarf contributes most of the near-ultraviolet flux we find an upper limit for the white dwarf effective temperature of $T_\mathrm{eff}\le16\,500$\,K.  The overplotted curve is for a white dwarf model at 16\,500 K. }
\label{fig:SED2}  
\end{center}
\end{figure}

Assuming that the near-ultraviolet emission of KIC~5608384 originates largely from the white dwarf provides an upper limit on its temperature. We adopt the white dwarf mass from the MCMC fit of the system parameters (Sect\,\ref{sec:mcmc}), $0.47\,M_\odot$, the distance of 367\,pc, and the mass-radius relation of \citep{holberg2006}, and find that $T_\mathrm{eff}\la16\,500$\,K is consistent with the observed near-ultraviolet fluxes (Fig.\,\ref{fig:SED2}). 

\begin{table}
\centering
\caption{UV photometry of KIC 5608384 (AB magnitudes)}
\begin{tabular}{lccc}
\hline
\hline
Band & Effective~$\lambda$ & Exposure & Magnitude \\
 & [\AA] & [sec] &  \\
\hline
\textit{Swift} uvw2 & 1191 & 384  & $18.50\pm0.06$\\
\textit{Swift} uvm2 & 2221 & 1163 & $18.55\pm0.06$\\
\textit{Swift} uvw1 & 2274 & 1150 & $18.09\pm0.05$\\
\textit{GALEX} NUV  & 2486 & 1283 & $18.59\pm0.02$\\
\hline
\label{tbl:uvmags}  
\end{tabular}
\end{table} 

\section{Inference of the K-star Radius} 
\label{sec:SED}

We determine the physical parameters of KIC 5608384 from the spectral-energy-distribution (SED). To this end, we used photometric magnitudes from Gaia ($G$, $G_{BP}$, and $G_{RP}$; \citealt{gaia18}), the Two-micron all-sky survey ($J$, $H$, and $K_S$; \citealt{cutri03}), the Wide-field Infrared Survey Explorer ($W1$ and $W2$; \citealt{cutri13}), the AAVSO Photometric All-Sky Survey ($B$, $V$, and $r'$; \citealt{henden16}), and the Carlsberg Meridian Catalogue ($r'$; \citealt{muinos14}). 

To determine $F_{\rm bol}$ and $T_{\rm eff}$, we simultaneously compared the SED and observed optical spectrum of KIC 5608384 (Sect.~\ref{sec:spectra}) to a grid of template spectra. The basic method is described in detail in \citet{mann16}. and briefly summarized here. For each template, we first reddened the spectrum by an estimated E(B-V), which is explored as a free parameter in the fit $[A(V)/E(B-V)$ assumed to be 3.1]. We then generated synthetic magnitudes using the relevant filter profiles and zero-points from Cohen et al. (2003), \citet{jarrett13}, \citet{mann15}, or \citet{apellaniz18}. The resulting (reddened) synthetic magnitudes were compared to the observed values, and the reddened spectrum was compared to the observed one. This comparison assumed 3\% errors in flux calibration in both the templates and KIC 5608384 spectrum, and accounts for errors in the filter zero-points. Gaps in the spectrum were replaced by BT-SETTL atmospheric models \citep{allard12}, which also provided an estimate of $T_{\rm eff}$. We calculated $F_{\rm bol}$ by integrating. 

When combined with the Gaia distance, this procedure yielded an estimate of $L_*$. We then joined $L_*$ and $T_{\rm eff}$ to derive $R_*$ from the Stefan-Boltzmann relation ($L_* = 4\pi \sigma R_*^2 T_{\rm{eff}}^4$). This process is repeated over all templates and $E(B-V)$ values, each time recording the values and goodness-of-fit ($\chi^2$). This process yielded $E(V-B )= 0.05 \pm 0.03$, $T_{\rm eff}$= $4299 \pm 80$ \,K,  $L_* = 0.174  \pm  0.08$, and $R_* = 0.754 \pm 0.04$. The errors account for imperfect selection of a template, but may underestimate uncertainties due to systematics in the atmospheric models.  These results are summarized in Table \ref{tbl:binary}.  

In these calculations we have neglected the contribution to the system light from the white dwarf and accretion disc/hot spot in quiescence.  We estimate that these do contribute $\sim$8\% of the bolometric flux and about $\sim$3\% of the light in the {\em Kepler} band. We conclude that this would affect the parameters given in Table \ref{tbl:binary} by less than or about the size of the cited uncertainties.  

\section{Long-Term Mass-Transfer Rate} 
\label{sec:lum}

The long-term average luminosity of white dwarfs, $L_{\rm eq}$, in CVs is set by compressional heating \citep{townsley2003} which depends on the long-term average accretion rate, $\dot M$, and the mass of the white dwarf.  Therefore, we can utilize the approximate $T_{\rm eff}$ estimated from the UV fluxes from KIC 5608384 (Sect.~\ref{sec:UV}) to infer the long-term $\dot M$ .  To this end, we make use Eqn.~(1) of \citet{townsley2009}
\begin{equation}
L_{\rm eq} \simeq 0.0057 \,L_\odot\, \langle \dot M \rangle_{-10} \left(M_{\rm wd}/M_\odot\right)^{0.4}
\label{eqn:LMdot}
\end{equation}
where $\langle \dot M \rangle_{-10}$ is the average mass-transfer rate in units of $10^{-10} M_\odot$ yr$^{-1}$.  For WD masses near $0.5 \,M_\odot$ we can approximate the radius-mass relation approximately by a power law to find:
\begin{equation}
R_{\rm wd} \simeq 0.0092 \, M_{\rm wd}^{-0.6} \, R_\odot$$
\label{eqn:Rwd}
\end{equation}
Combining this relation with Eqn.~(\ref{eqn:LMdot}) we can relate the long-term $T_{\rm eff}$ with the long-term $\dot M$:
\begin{equation}
T_{\rm eff} \simeq 12,800 \, \langle \dot M \rangle_{-10} \left(M_{\rm wd}/M_\odot\right)^{0.4} \, {\rm K}
\label{eqn:TMdot}
\end{equation} 

If we now take the WD temperature to be $\lesssim 16,500$ K we can utilize Eqn.~(\ref{eqn:TMdot}) to deduce a secular mean accretion rate of 
$$\langle \dot{M} \rangle \lesssim 3 \times10^{-10}\,M_\odot\,\mathrm{yr}^{-1}$$
averaged over $\sim10^5$\,yr. This rate is about a factor of twenty below the secular mean accretion rate predicted by the evolutionary sequence described in Sect.\,\ref{sec:evolve2}. 

KIC~5608384 is not alone in having an accretion rate that is far below the predictions for its orbital period: HS\,0218+3229 has very similar binary parameters, including an evolved donor star \citep{rodriguez-gil09}, and analysing \textit{HST} ultraviolet spectroscopy of this system, \citet{pala2017} found a white dwarf temperature of 17\,990\,K, corresponding to an accretion rate of $\dot{M}\simeq4\times10^{-10}M_\odot\,\mathrm{yr}^{-1}$. Again, this rate is more than an order of magnitude below the predictions, raising the question if the standard prescriptions underlying these evolutionary calculations are correct.

\section{Deducing the System Parameters} 
\label{sec:mcmc}

In this section we describe the five constraints we use to help infer the binary system parameters.  We utilize an MCMC approach (see, e.g., \citealt{ford05}; \citealt{madhu09}, and references therein).  The constraints are: (1) the K velocity from the radial velocity curve (Sect.~\ref{sec:spectra}); (2) the ELV amplitude (Sect.~\ref{sec:Kepler}); (3) the Roche-lobe filling condition; (4) the inferred radius of the K star (Sect.~\ref{sec:SED}); and (5) the existence of partial eclipses. From these, we would like to deduce the mass of the K star, $M_{\rm K}$, the mass of the white dwarf, $M_{\rm wd}$, and the orbital inclination angle, $i$. Specifically, the MCMC code chooses the two constituent masses, the K-star radius, and the orbital inclination angle, and then tests via the $\chi^2$ statistic the agreement with the input constraints.

In somewhat more detail, what follows is a description of specific constraints. \\
\noindent
{\em (i) Mass Function from RV curve}:
The constituent masses and inclination angle for each link of the MCMC chain yield a predicted value of the mass function of the system from: $f(M)=M_{\rm wd}^3\sin^3i/(M_{\rm wd}+M_{\rm K})^2$.  This value is then tested, via $\chi ^2$, against the value determined from the radial velocity curve for the K star and the known orbital period, $f(M) = 0.1238 \pm 0.007 \, M_\odot$.  
\vspace{3pt}

\begin{figure}
\begin{center}
\includegraphics[width=1.01 \columnwidth]{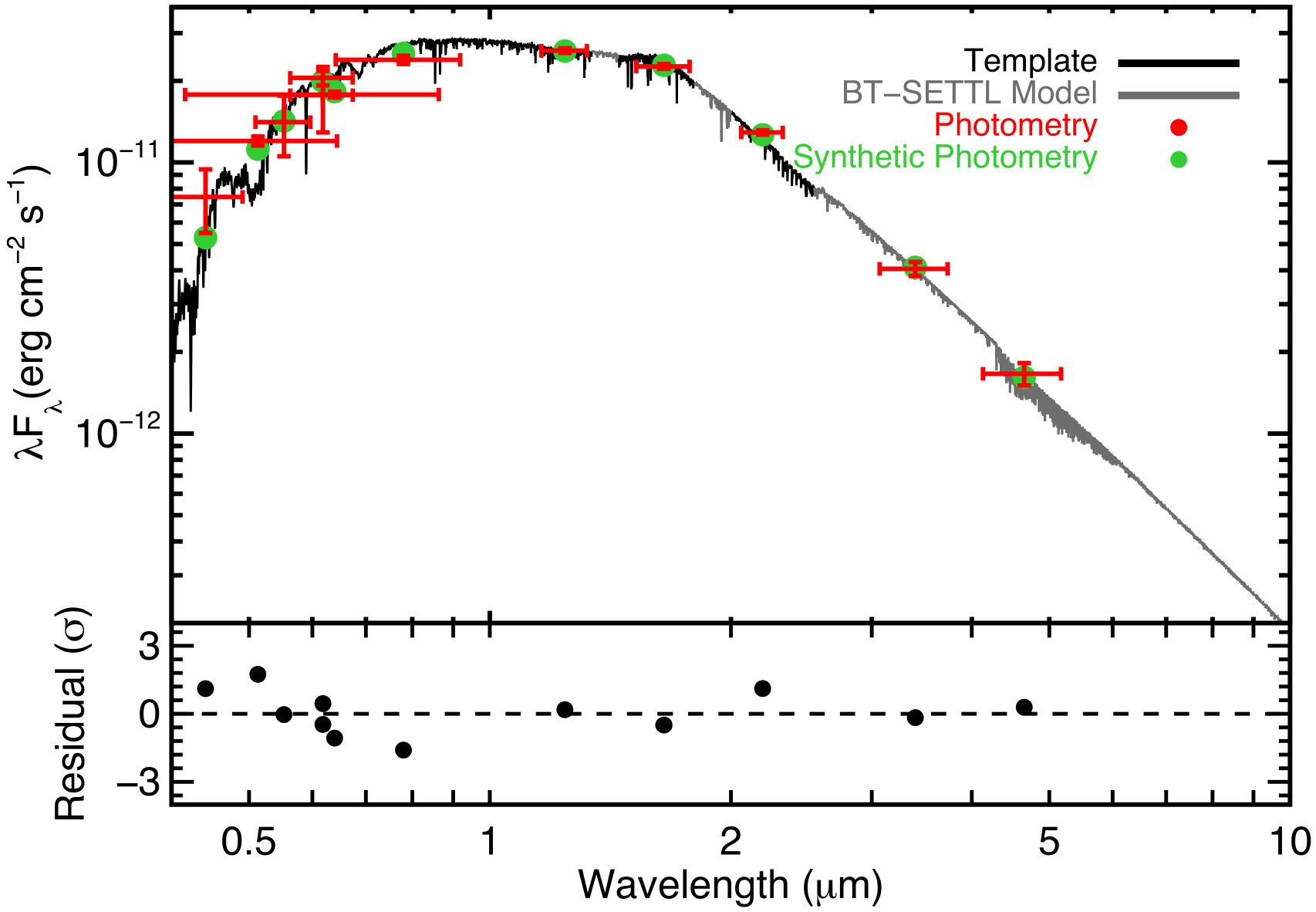} 
\caption{SED of KIC 5608384, with the optical defined by the Gaia $G_\mathrm{BP}$ and $G_\mathrm{RP}$, and the infrared by the 2MASS and WISE magnitudes. Observed magnitudes are shown as error bars (in red), with vertical errors corresponding to the observed magnitude uncertainties (including stellar variability) and the horizontal errors corresponding to the width of the filter. Synthetic magnitudes from the best-fit template spectrum (black) and BT-SETTL model (grey) are shown as large fainter (green) circles. (See Sect.~\ref{sec:SED} for details.)}
\label{fig:SED} 
\end{center}
\end{figure}

\begin{figure}
\begin{center}
\includegraphics[width=0.8 \columnwidth]{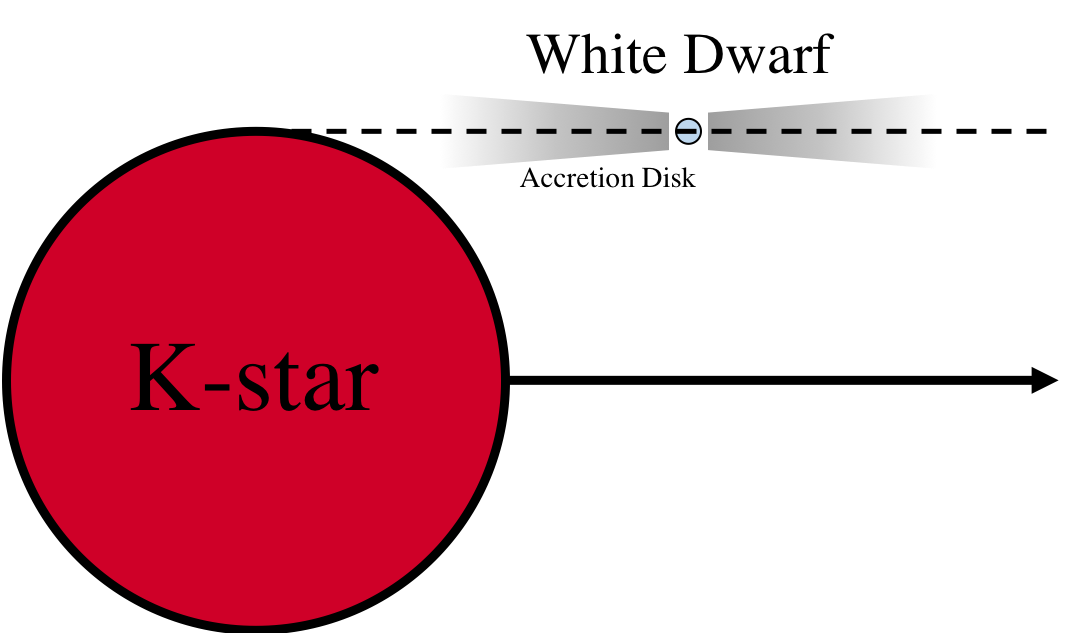} 
\caption{Schematic sketch of the likely eclipse geometry.  At least partial eclipses by the K star are observed of the H$\alpha$ emission line region in quiescence and of the accretion disk, and possibly the white dwarf, in the {\em Kepler} band during the CV outburst.}
\label{fig:diagram} 
\end{center}
\end{figure}

\noindent
{\em (ii) Amplitude of ellipsoidal light variations}:
The amplitude of the ELVs in a binary can be represented analytically (see \citealt{kopal59}) approximately by an expression of the form: $A_{\rm ELV} \approx C (M_{\rm wd}/M_{\rm K})(R_{\rm K}/a)^3 \sin^2 i$, where $a$ is the orbital separation, and $C$ is a dimensionless constant of order unity. Since the K star is assumed to be filling its Roche lobe, $(R_{\rm K}/a)$ is a function only of the mass ratio, $q \equiv M_{\rm K}/M_{\rm wd}$, and we can therefore write the entire expression for $A_{\rm ELV}$ (modulo the term involving the inclination) as a function of $q$.  We have used the code {\tt LIGHTCURVEFACTORY} \citep{borko13,rappaport17,borkovits18} to generate ellipsoidal light variations in generic binaries with a wide range of $q$ values, but all for a fixed $T_{\rm eff} = 4400$ K\footnote{This value of $T_{\rm eff}$, used to model the ELV amplitudes, $A_{\rm ELV}$, is sufficiently close to our final value for $T_{\rm eff}$ so that any differences in the resultant $A_{\rm ELV}$ are negligible compared to other uncertainties in the analysis.} and a Roche-lobe-filling star.  We then used these results to generate a fitting formula of the following form: $A_{\rm ELV} \simeq (0.154689 - 0.067022 \,q + 0.019436 \,q^2 - 0.001956 \,q^3) \times \sin ^2 i$. 

\begin{figure*}
\begin{center}
\includegraphics[width=0.45 \textwidth]{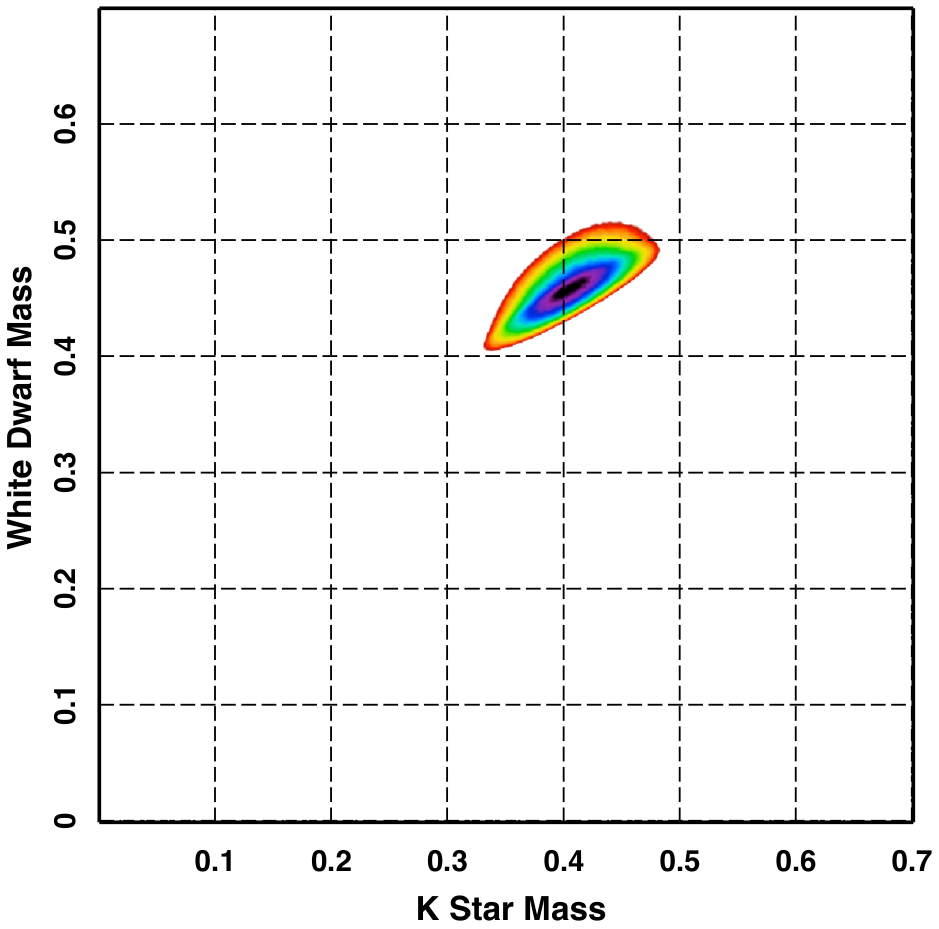} \hglue0.1cm
\includegraphics[width=0.45 \textwidth]{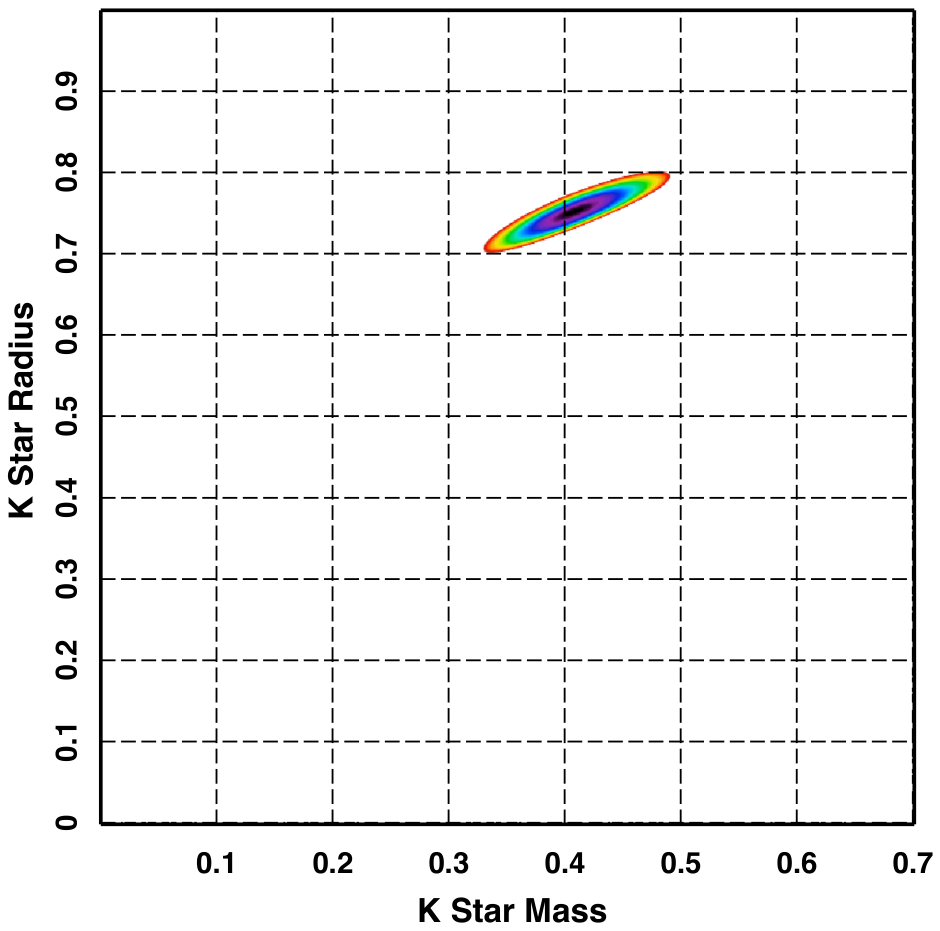} 
\caption{MCMC correlation plots for the white dwarf mass, $M_{\rm wd}$ and K-star mass, $M_K$, and radius, $R_K$, in the KIC 5608384 system.  These are deduced from the radial velocity curve (see Sect.~\ref{sec:spectra}), the inferred radius for the K star from SED fitting (see Sect.~\ref{sec:SED}), the Roche-lobe filling condition, and the amplitude of the ELVs (see Sect.~\ref{sec:mcmc} for details). }
\label{fig:MCMC} 
\end{center}
\end{figure*}

The constituent masses and inclination angle for each link of the MCMC chain then yield a predicted value for the amplitude of the ELV which is then compared to the measured ELV amplitude of $0.118 \pm 0.018$ (Sect.\,\ref{sec:Kepler}). The {\em model}  ELV amplitudes are somewhat arbitrarily taken to be uncertain by $\approx \pm 15\%$ due to the estimated uncertainties in the model calculations.
\vspace{1pt}

\noindent
{\em (iii) Roche lobe filling constraint}:
Because there is manifestly mass transfer taking place in KIC 5608384, we assume that the radius of the K star is equal to the radius of its Roche lobe.  In turn, the Roche lobe radius for the K star is given by the analytic expression: $R_{\rm L} \simeq 0.49 \, a \, q^{2/3}/[0.6q^{2/3}+\ln(1+q^{1/3})]$ \citep{eggleton83}.  We then compare this radius with $R_{\rm K}$ of the current MCMC link, and assign a somewhat arbitrary `uncertainty' of 2\% in matching the Roche lobe while computing the contribution to $\chi^2$.
\vspace{3pt}

\noindent
{\em (iv) Inclination angle constraint}:
Because at least partial eclipses of the accretion disk are visible (both in quiescence and outburst; see Figs.~\ref{fig:lightcurve}, \ref{fig:fold}, and \ref{fig:trailed}), we impose a lower bound to the value of the inclination angle. At each link of the MCMC, we set this lower bound to be $i_{\rm min}=\cos^{-1}(R_{\rm K}/a)$, the inclination at which the limb of the K star passes through the center of the white dwarf in projection (see Fig.~\ref{fig:diagram}). After repeated tests, we concluded that this constraint does not significantly affect the inclination angle distribution already determined from the other constraints.

For every given set of MCMC parameters, we compute the total $\chi^2$ from the fit to the constraints. A decision about whether to accept the particular set of parameters or make a new selection is made based on the standard Metropolis-Hastings jump conditions (see, e.g., \citealt{ford05}; \citealt{madhu09}, and references therein).  We typically run $10^7$ links per chain, and repeat this a dozen times. From the outputs of the MCMC code we derive the system parameters $M_{\rm K}$, $M_{\rm wd}$, $R_{\rm K}$ and the inclination angle $i$.

\begin{figure}
\begin{center}
\includegraphics[width=1.00 \columnwidth]{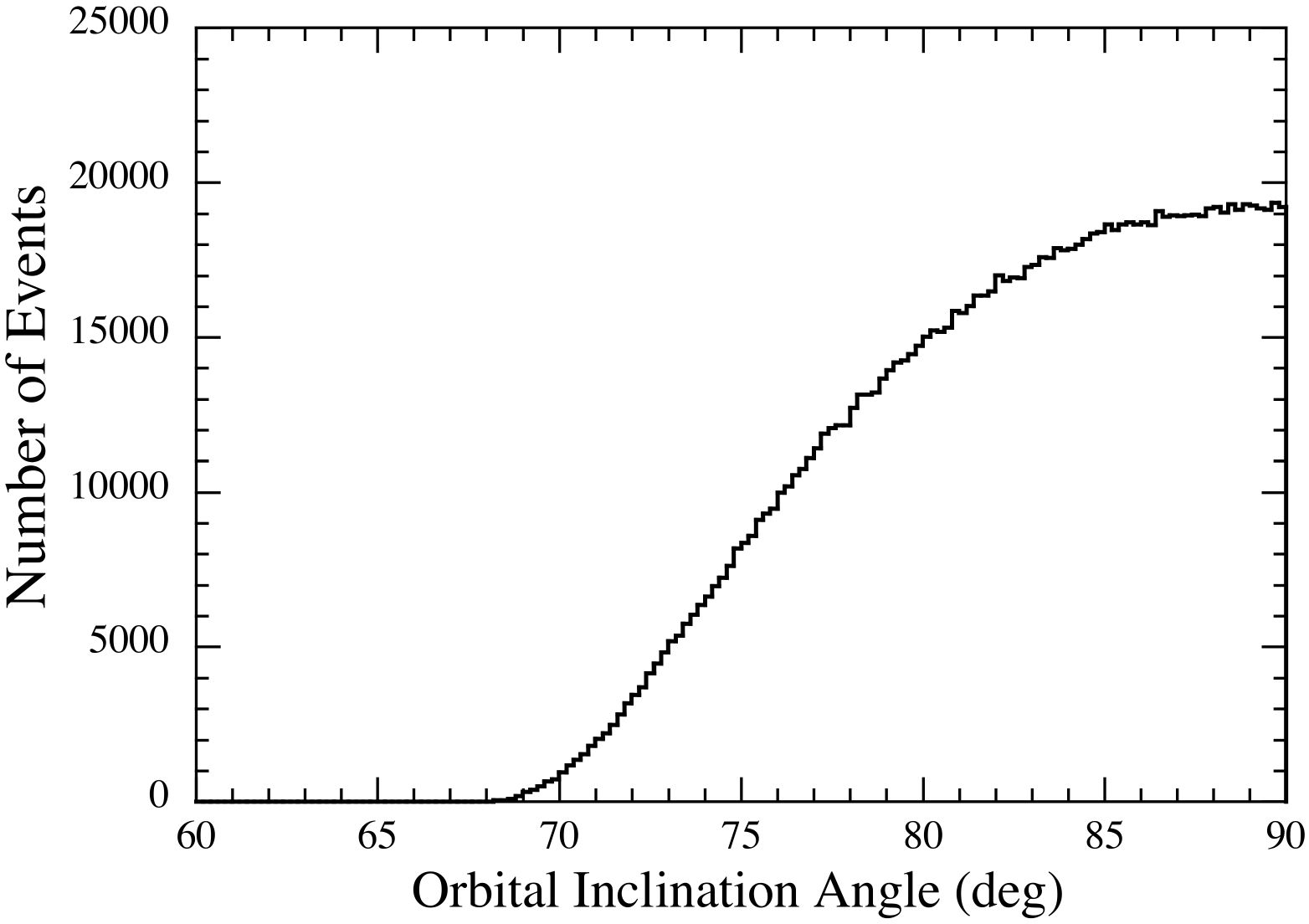} 
\caption{Monte Carlo probability density plot for the inclination angle of the KIC 5608384 system based on the same input information described in Fig.~\ref{fig:MCMC}.}
\label{fig:incl} 
\end{center}
\end{figure}

The results of the MCMC analysis are given as correlation plots in Fig.~\ref{fig:MCMC}.  The left panel shows the correlation between $M_{\rm wd}$ and $M_{\rm K}$, while the right panel displays the correlation between the radius and mass of the K star.  Figure \ref{fig:incl} shows separately the distribution of allowed inclination angles. The best-fit values and the 1-$\sigma$ uncertainties are summarized in Table \ref{tbl:binary}.

\begin{table}
\centering
\caption{Summary Properties of the KIC 5608384 Binary}
\begin{tabular}{lc}
\hline
\hline
Parameter & Value \\
\hline
$P_{\rm orb}$$^a$ [day] & 0.3641106\,(4) \\
$P_{\rm orb}$$^a$ [hr] & 8.73865\,(1) \\
$a$$^b$ [$R_\odot$] & $2.04 \pm 0.04$ \\
$T_0$$^a$ [BJD] & 2452920.1112\,(70) \\
$K_{\rm K}$$^c$ [km s$^{-1}$] & $148.6 \pm 2.8$ \\
$K_{\rm K}$$^b$ [km s$^{-1}$] & $150.2 \pm 2.5$ \\  
$K_{\rm wd}$$^b$ [km s$^{-1}$] & $133.5 \pm 5.0$ \\  
$M_{\rm K}$$^b$ [$M_\odot$] & $0.406 \pm 0.028$ \\
$R_{\rm K}$$^d$ [$R_\odot$] & $0.754 \pm 0.040$ \\
$T_{\rm K}$$^d$ [K] & $4300 \pm 80$ K \\
$F_{\rm bol,K}$$^d$ [(ergs cm$^{-2}$ s$^{-1}$) & $(4.20\pm 0.13)\times 10^{-11}$ \\
$L_{\rm bol, K}$$^d$ [$L_\odot$] & $0.174\pm 0.008$ \\
$M_{\rm wd}$$^b$ [$M_\odot$] & $0.458 \pm 0.019$ \\
$R_{\rm wd}$$^e$ [$R_\odot$] & $0.0147 \pm 0.0004$ \\
$T_{\rm wd}$$^f$ [K] & $\simeq 16500$ \\
$L_{\rm bol, wd}$$^{e,f}$ [$L_\odot$] & $\simeq 0.015$ \\
$\langle \dot M \rangle$$^g$ [$M_\odot$ yr$^{-1}$] & $\sim$$3 \times 10^{-10}$ \\
$\langle \dot M \rangle$$^h$ [$M_\odot$ yr$^{-1}$] & $6.5 \times 10^{-9}$ \\
$E(B-V)$$^i$ & $0.05\pm 0.03$ \\
$i$$^b$ [deg] & $\gtrsim 70^\circ$ \\
$f(M)$$^c$ [$M_\odot$] & $0.1238 \pm 0.0071$ \\
$\tau_{{CV}}$$^j$ (${\rm age}$) [Gyr] & $\gtrsim 7.5$ \\
$M_{\rm K,post-CE}$$^j$ [$M_\odot$] & $1.07 \pm 0.04$ \\
$P_{\rm bin, post-CE}$$^j$ [days] & $3.4 \pm 0.2$ \\
$P_{\rm ZACV}$$^j$ [days] & $0.64 \pm 0.03$ \\

\hline
\label{tbl:binary}  
\end{tabular}

(a) Based on the {\em Kepler} photometry. (b) Derived from the MCMC fitted parameters. (c) From the RV analysis. (d) Based on the SED analysis presented in Sect.~\ref{sec:SED}. (e) See Eqn.~(\ref{eqn:Rwd}).  (f) Based on the SWIFT and Galex observations (see Sect.~\ref{sec:UV}). (g) Based on the current $T_{\rm eff}$ of the WD and the inferred compressional heating by the accreted material; see Sect.~\ref{sec:lum}.  (h) This value of $\dot M$ is taken from the {\tt MESA} calculated values. (i) From the SED analysis (see Sect.~\ref{sec:SED}); this value of $E(B-V)$ was corroborated, after the fact, at the on-line tool ``3D Dust Mapping with Pan-STARRS 1'' \url{http://argonaut.skymaps.info/} \citep{green18} with a value of $0.04 \pm 0.02$.  (j) Values inferred from {\tt MESA} evolutionary tracks.
\end{table}  

\section{The Origin of KIC 5608384} 
\label{sec:origin}

\subsection{Formation Scenarios}
\label{sec:evolve}

The canonical model for the formation and evolution of cataclysmic variables has been discussed extensively in the literature (see, e.g., \citealt{rappaport83}; \citealt{patterson84}; \citealt{warner95}; \citealt{howell01}; \citealt{knigge11}; \citealt{goliasch15}; \citealt{kalomeni16}, and references therein).  While the more massive primary is evolving up the red giant or asymptotic giant branch it overflows its Roche lobe (i.e., RLOF) leading to  dynamically unstable mass transfer.  Assuming that a merger is avoided, the ensuing phase of common envelope (CE) evolution produces a white dwarf in a tight orbit with the nearly pristine secondary star  (see, e.g., \citealt{paczynski76}; \citealt{taam78}; \citealt{webbink84}; \citealt{taam92}; \citealt{pfahl03}; \citealt{ivanova13}, and references therein). 

Either as the result of nuclear evolution of the secondary or the orbital separation shrinking due to angular momentum losses (e.g., magnetic braking; see \citealt{rappaport83}), the secondary continuously fills a larger fraction of the volume of its Roche lobe.  If the secondary eventually overflows its Roche lobe (i.e., ZACV) and if mass transfer is dynamically stable{\footnote{This assumes that `latent' dynamical instabilities do not occur (see \citealt{goliasch15}).}}, then, depending on the initial conditions, there are three possible evolutionary outcomes: (i) if the donor is not chemically evolved, the CV will follow the `classic track' by evolving from longer orbital periods ($\lesssim 10$ hours) to shorter ones before reaching a minimum period of $\simeq$72-75 minutes\footnote{We note that the {\em observed} minimum orbital period of CVs is about 80 minutes.} and then evolving back to longer periods as the secondary attains a mass of $\lesssim 0.05 M_\odot$; (ii) if the donor is sufficiently evolved, the ZACV may evolve to ultrashort periods of $\gtrsim 5$ minutes before the orbital period increases (these would be classified as AM CVn binaries); and, (iii) if the donor is even more evolved, the orbital period of the binary may increase enormously (up to $\lesssim 1000$ hours) as the secondary (donor star) becomes a giant.   The delineation between cases (ii) and (iii) is known as the `bifurcation' (see \citealt{PS}; and, \citealt{kalomeni16}, for an extensive review).  Based on the inferred present-day properties of KIC 5608384, we believe that the initial conditions place it close to the bifurcation.

\begin{figure}
\includegraphics[width=1.00 \columnwidth]{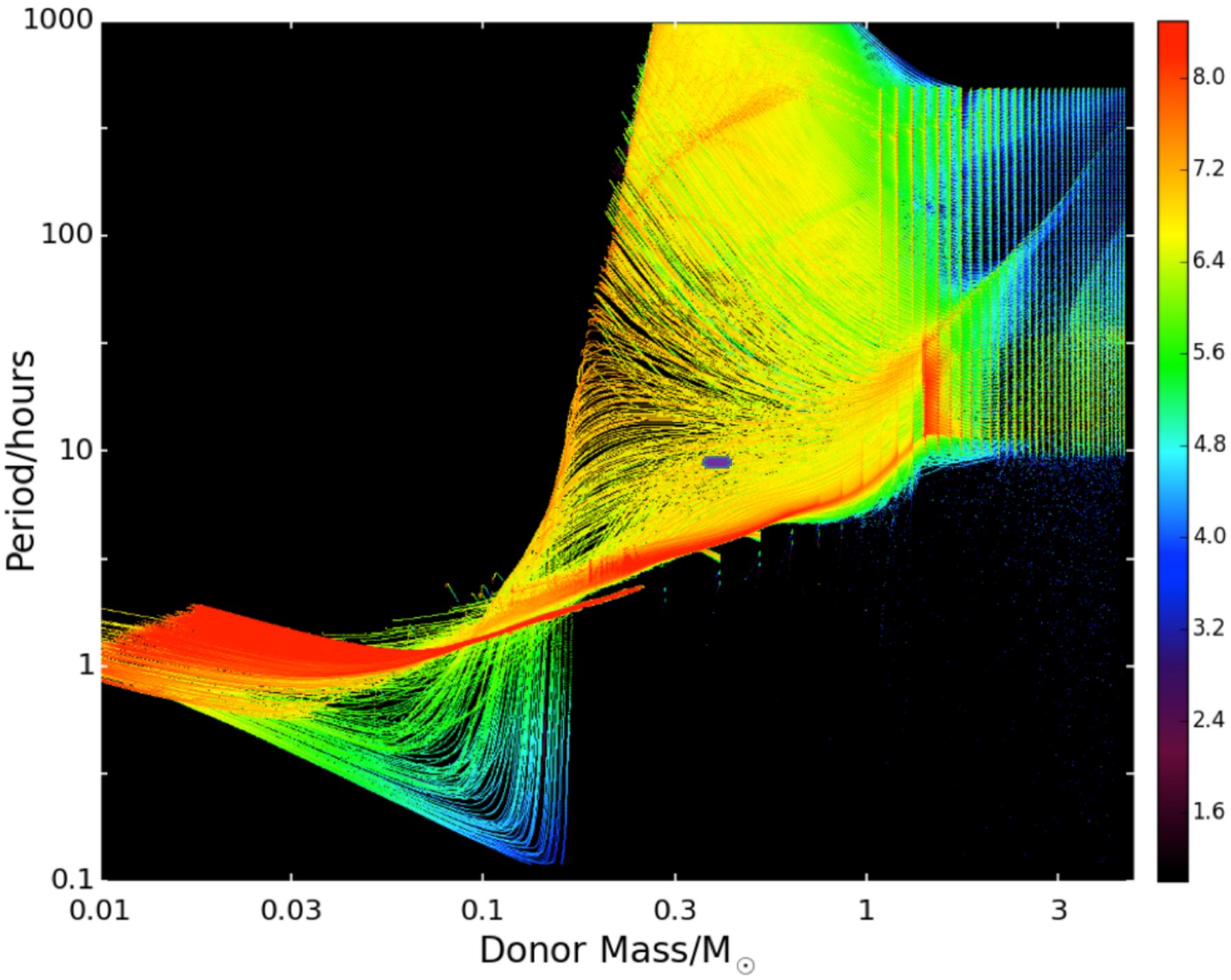}
\caption{40,000 model CV evolution tracks in the orbital period, $P$, donor mass, $M_{\rm K}$ plane (from \citealt{kalomeni16}). The heavy elongated purple marker indicates the location of KIC 5608384. The evolution tracks all start in the vertically striped region in the upper right with $10 \lesssim P \lesssim 400$ hrs and $1 \lesssim M \lesssim 4 \, M_\odot$. Color indicates the logarithm of the evolutionary dwell time.  Depending on exactly where the pre-CV system starts within this region, it can evolve along a conventional CV track (red region), up the giant branch (yellow region), and to ultrashort periods (green and blue regions). KIC 508384 sits in a transition region with systems starting near the so-called bifurcation, and will likely evolve to ultrashort periods as AM CVns.}
\label{fig:P_M_diagram} 
\end{figure}

\begin{figure}
\includegraphics[width=0.95 \columnwidth]{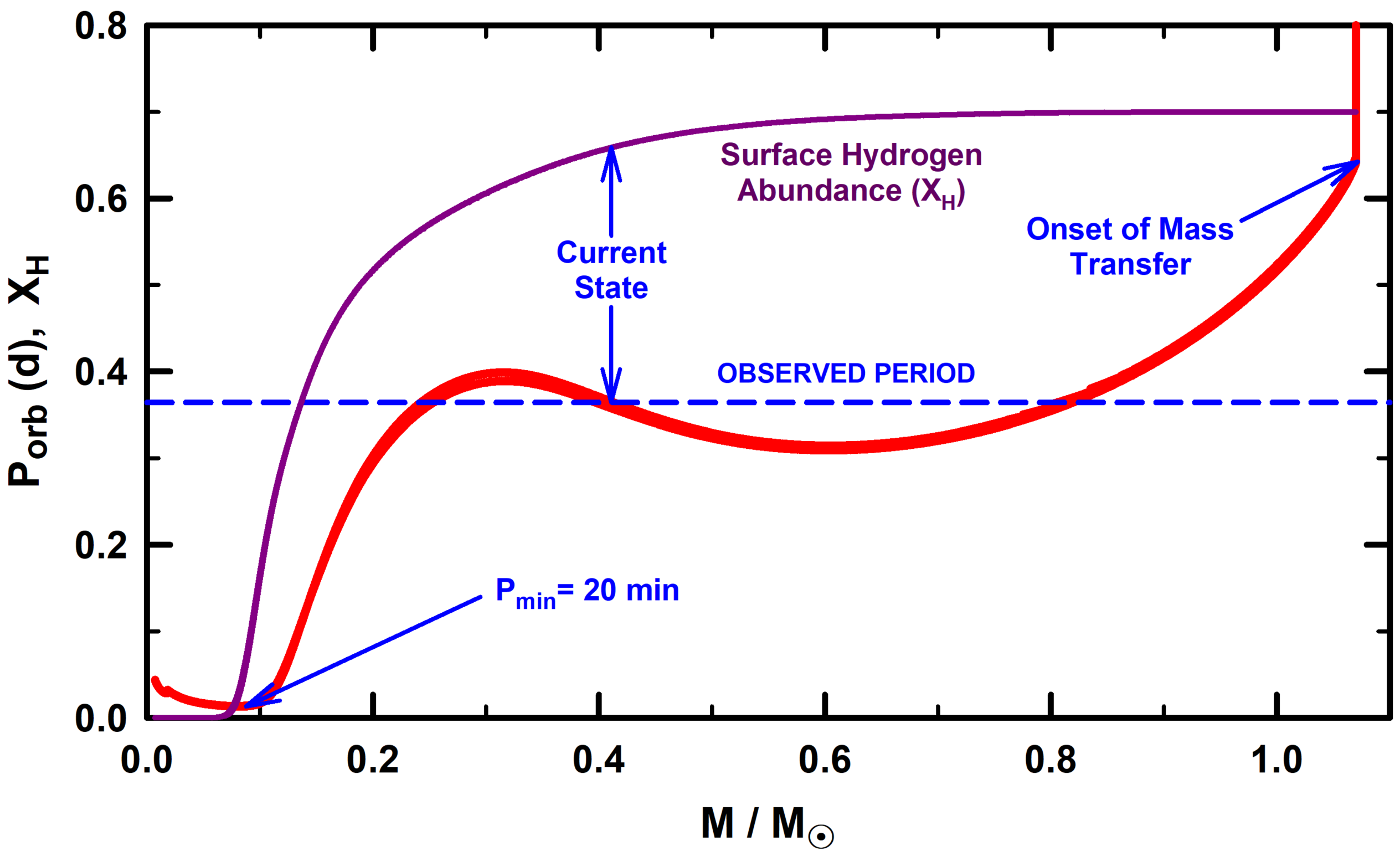} \hglue-0.04cm  
\includegraphics[width=0.99 \columnwidth]{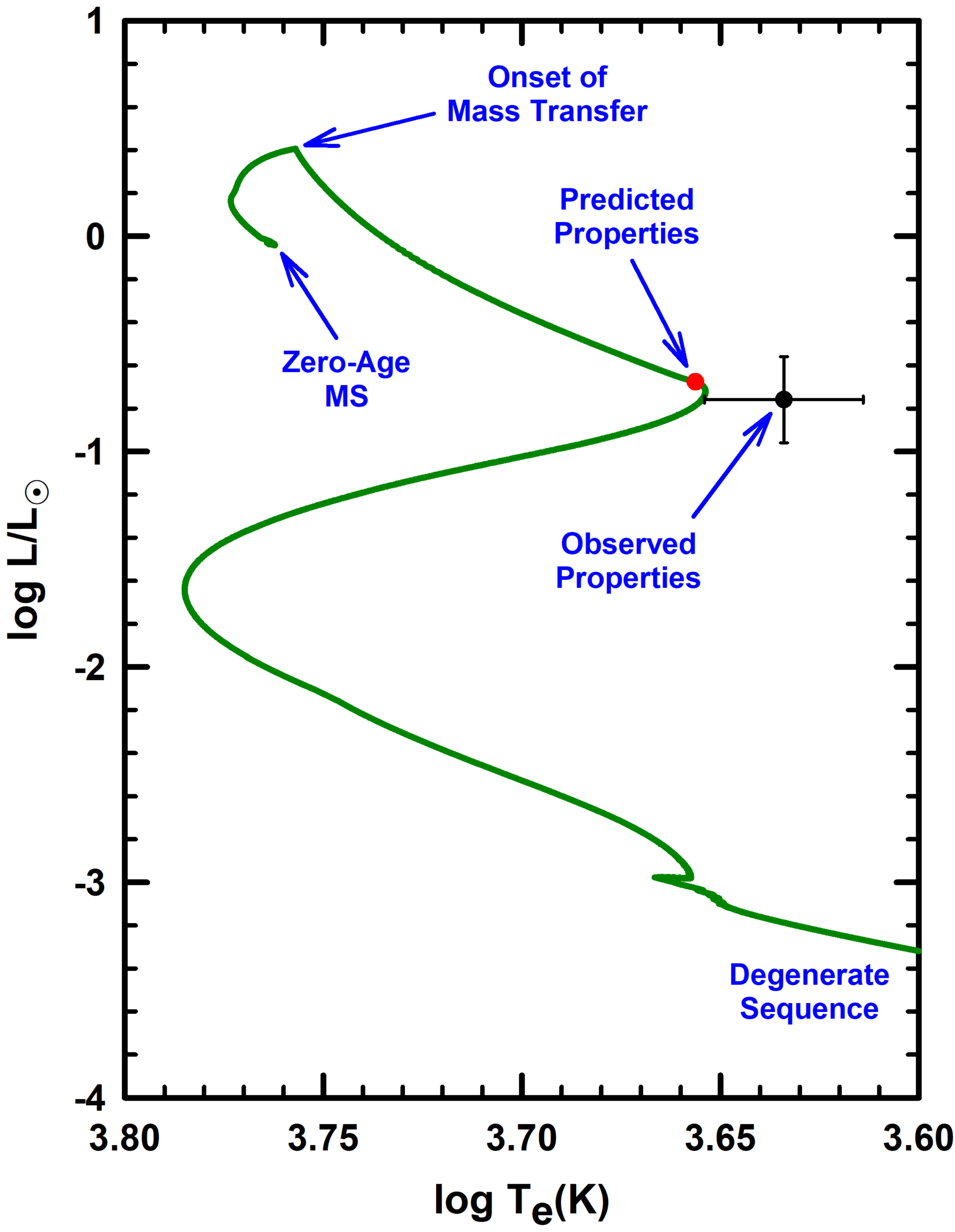} 
\caption{Illustrative CV evolution tracks, starting from a ZACV, that roughly match the current status of KIC 5608384 (run with the {\tt MESA} evolution code; \citealt{paxton13,paxton15,kalomeni16}).  Top panel shows the evolution of $P$ and $X$ (the surface hydrogen mass fraction) as functions of the donor-star mass, $M$, for three tracks (initial periods of 3.41 to 3.43 days; only the finite thickness of the curves reveals the presence of three tracks.  Note how the systems reach a local minimum in the orbital period of $\simeq$7.5 hrs, then rise to 9.5 hours, before finally plummeting to an ultrashort period of about 20 minutes.  Bottom panel shows the corresponding evolution in the HR diagram. Note that the theoretical models predict temperatures and luminosities that are slightly higher than the observationally inferred ones.}
\label{fig:tracks} 
\end{figure}

To explore which ZACVs can evolve to the current state of KIC 5608384, we have examined some 40,000 CV evolutionary tracks produced with the {\tt MESA} stellar evolution code \citep{paxton13,paxton15} for the CV study by \citet{kalomeni16}.  In Fig.~\ref{fig:P_M_diagram} (taken from \citealt{kalomeni16}) we reproduce the ensemble of CV evolution tracks in the orbital period-donor mass ($P-M$) plane in order to constrain the properties of the ZACV.  We have superposed the currently measured properties (with error bars) of KIC 5608384 on that plot.  As is readily apparent, KIC 5608384 does not lie along the conventional CV tracks.  These are seen as red regions in Fig.~\ref{fig:P_M_diagram} and correspond to the high probability cases (i.e., the `canonical' model).  We also see that its position does not overlay tracks for CVs that evolve to much higher periods (i.e., ascending the giant branch).  Although the properties of KIC 5608384 place it close to the bifurcation region, the evolutionary tracks show that possible prototype systems will ultimately evolve to ultrashort periods of $\sim$5-40 minutes (the AM CVn channel referred to as case (ii) above).

\subsection{Hot subdwarf phase?}
\label{sec:sdB}

Before trying to identify possible ZACV models that will evolve to produce the currently  observed properties of KIC 5608384, we discuss the intriguing possibility that the accretor in this system actually evolved through a hot subdwarf phase before cooling to become the currently observed $\simeq 16500$ K WD companion.  

Hot subdwarfs were discovered nearly 70 years ago and were observed to have very high temperatures (e.g., sdB stars have temperatures of $\approx 30000$K) but have luminosities that place them well below their main-sequence counterparts on the HR diagram (see \citealt{Heber}, and references therein).  It is suspected that sdB and sdO subdwarfs are so numerous that they are responsible for the `UV-upturn' observed in many elliptical galaxies. 

According to \citet{Han}, hot subdwarfs with mass distributions that are sharply peaked at $\simeq 0.46$ to  $0.47$~M$_\odot$ can be formed in close binaries after a single phase of CE evolution.  Based on the inferred properties of KIC 5608384, it is quite possible that the binary evolved via this channel.  Taking the preferred model parameters deduced by \citealt{Han} (their model set [2]\footnote{For this specific set, they made conservative assumptions with respect to the critical mass ratio of the components that would lead to CE ejection (i.e., $q_{\rm crit}=1.5$) and further assumed relatively high efficiencies for CE removal and thermal energy feedback ($\alpha_{CE}=\alpha_{th}=0.75$, respectively). Note that other choices of these parameters can also produce similar ZACVs containing hot subdwarfs.}), they show that close binaries containing hot subdwarfs with companion masses of  $\simeq 1.1$~M$_\odot$ and orbital periods of approximately 3.5 days can be formed (see, e.g., their Figure 15).  These are the initial conditions needed to reproduce the properties of KIC 5608384 (see section~\ref{sec:evolve2} for a detailed discussion).  Based on their models, we further conclude that the mass of the subdwarf's progenitor was $\approx 1.7 \, M_\odot$  and that the primordial binary for such a mass had an orbital period of $\approx 300$ days.  If the limiting value of $q_{\rm crit}$ is relaxed, then it is possible for the subdwarf's progenitor to be a few tenths of a solar mass smaller (but at the cost of increasing the age of the binary to possibly unacceptably high values).  Thus we believe that the higher-mass value of the progenitor is more likely.

After the CE phase, the core of the progenitor evolves to the sdB stage and begins converting helium (He) into carbon over a period of $\sim100$ Myr.  The sdO phase ensues shortly thereafter wherein oxygen is created as a result of the $\alpha$-channel capture. An example of a typical evolution of a subdwarf from the sdB to the sdO phase, and then to a cold WD is shown in \citet{Zhou}.  The canonical mass of these subdwarf remnants is thought to be about  $0.47 \, M_\odot$.  Given that the mass we infer for the compact companion in KIC 5608384 ($0.46 \pm 0.02 \, M_\odot$),  is extremely close to the expected mass of WDs that underwent a hot subdwarf phase, it is quite plausible that the WD was a hot subdwarf.  Alternatively, it is also possible that the primary underwent a CE evolution while it was either: (i) near the tip of the Red Giant Branch (RGB) thereby producing a He WD; or (ii) was near the base of the Asymptotic Giant Branch (AGB) thereby producing a CO WD.  However, because He cores at the tip of the RGB can only have a maximum mass of $\simeq 0.46 \, M_\odot$, it seems unlikely that the primary would have evolved so close to the tip of the RGB.  With respect to the second possibility, CO WDs have masses of $\gtrsim 0.52 \, M_\odot$.  Thus, we believe that the WD was likely a hot subdwarf that emerged from the CE channel as originally described by \citet{Han}.

\subsection{Evolution from the ZACV}
\label{sec:evolve2}

In order to infer the evolutionary properties of KIC 5608384 we generated a grid of CV evolution models using version 10108 of {\tt MESA}.  Our calculations are similar to those carried out by \citet{kalomeni16}.  We adopted a solar metallicity based on the system's location in the Galaxy. To determine the properties of the ZACV, we have taken the mass of the WD to be $0.47 \, M_\odot$ and further set the \citet{TaurisvdH} mass and angular momentum loss parameters such that $\alpha = 0$ and $\beta = 1$.  Imposing these constraints is equivalent to assuming that all of the mass lost from the donor (secondary) is temporarily accreted by the WD and then subsequently lost from the binary as a result of classical nova explosions.  Thus we assume that all of the mass that is initially accreted by the WD is completely lost from the binary and carries away the specific angular of the WD (i.e., `fast Jeans' mode').  This assumption is always checked {\sl a posteriori} by examining whether the mass transfer rate was ever sufficiently high to allow for a `supersoft X-ray source' phase that allowed for the non-explosive burning of hydrogen on the surface of the accretor (see, e.g., \citealt{diSN}).  Orbital angular momentum dissipation was calculated based on the torques associated with gravitational radiation and magnetic braking as described in \citet{kalomeni16}, with the magnetic braking index set equal to 3.  It should be noted that the magnetic braking formula (Verbunt-Zwaan law) was inferred from observations of low-mass main-sequence stars and thus must sometimes be extrapolated to stars that are either evolved, of very low-mass, or rapidly rotating. Thus a precise calculation of the effects of magnetic braking remains problematic.  Other angular momentum loss mechanisms have been proposed such as consequential angular momentum losses (CAML; see \citealt{schreiberetal16}) but have not been investigated in this paper.

The grid of {\tt MESA} models was created by treating the mass of the secondary (donor) and the orbital period after the CE phase as free parameters. The evolutionary tracks of the models were followed until the present day.  We were then able to identify the volume of (initial-condition) phase space that admitted models which reasonably match the currently observed properties of  KIC 5608384.  We found that with suitable adjustments for the post-CE binary orbital period (about 3.5 days), the initial secondary mass could have had values of $1.03 \lesssim  M_{2}/M_{\odot} \lesssim 1.12$.  This range of values ensured that the mass-transfer rates would be low enough to guarantee thermonuclear runaways (i.e., classical novae) and that the age of the binary (in most cases around 8 Gyr) would not be likely to exceed the limiting age of the disk of the Galaxy ($9 \pm 2$ Gyr; \citealt{soderblom}). 

An illustrative example of the evolution of three models all having a $1.07 \, M_{\odot}$ donor but for three different post-CE orbital periods is shown in Fig.~\ref{fig:tracks}.  The top panel shows the evolution of the orbital period, $P$, and $X$ (the surface hydrogen mass fraction) as functions of the donor-star mass, $M$.  Note that although the initial period for each model (post-CE) was about 3.4 days, the donor starts losing mass (ZACV phase) when the binary has attained an orbital period of about 15 hours.  Mass transfer initially proceeds at a reasonably rapid rate (i.e., on the thermal timescale of the donor) before declining to a rate of $\approx 6.5 \times 10^{-9} M_{\odot} \, {\rm yr}^{-1}$ for the present-day CV.  During the more rapid phase of mass transfer, the orbital period reaches a local minimum of $\simeq$7.5 hrs as the mass ratio approaches unity ($M \approx 0.6 \, M_\odot$) before increasing back to the observed 8.7 hours when the mass of the donor has decreased to $\simeq 0.41 \,M_{\odot}$.  The subsequent orbital evolution will be discussed in Sect.~ \ref{sec:PDCV}. 

The predicted accretion rate of KIC 5608384 is much larger than that estimated from its physical properties 
in Sect.~\ref{sec:lum} ($\simeq 3 \times 10^{-10} \, M_\odot {\rm yr}^{-1}$). We note that the observations of CVs 
with donors that underwent thermal-timescale mass loss suggest that these systems evolve at nearly constant 
accretion rates, $\simeq 0.7-2 \times 10^{-10} \, M_\odot {\rm yr}^{-1}$, from orbital periods of $\simeq 7$\,h down to 
$\simeq 1$\,h (see Fig.~2 in \citealt{toloza19}), and KIC 5608384 closely follows this trend. Whereas this finding is 
still based on only a handful of systems, it may suggest that the models for this evolution channel are not fully realistic.  
If, on the other hand, the fault is not with the models, there could be a selection effect at work, i.e., we might not recognise 
post-thermal timescale CVs as such if they have very high accretion rates because we would not detect the 
white dwarf.

Also plotted in the upper panel of the figure is the surface hydrogen abundance of the donor as a function of the donor's mass.  According to our models, the surface of KIC 5608384 should only be very slightly depleted of hydrogen relative to the primordial abundance and thus could not be detected observationally.  Nonetheless, our models require that the deep interior of the secondary must have been substantially evolved before mass loss started, thereby placing the binary close to  the `bifurcation' boundary.  This is demonstrated in the lower panel of Fig.~\ref{fig:tracks} which shows the evolution of the secondary (donor) in the HR diagram.  It indicates that the donor reached the Terminal Age Main Sequence (TAMS) and evolved as a subgiant before mass transfer commenced (ZACV).  The secondary continued to undergo nuclear evolution leading to lower surface temperatures while simultaneously experiencing thermal-timescale mass transfer.  According to the track followed by the donor, we expect that KIC 5608384 should be very close to this local extremum in temperature.  Because of the substantial amount of mass lost from the donor, we predict that the donor will subsequently be forced to follow a `horizontal branch' type of evolution that is typically seen as stars approach the degenerate sequence and become WDs.  However, unlike that scenario, the hydrogen-depleted core of the donor will not be sufficiently massive to allow it to form a hydrogen-exhausted He WD (due to the Sch{\"o}nberg-Chandrasekhar limit).   Instead, the secondary will become fully convective and the helium within the core will be mixed with the hydrogen-rich layer near the surface.  This ultimately causes the secondary to have a very hydrogen-depleted surface abundance (as can be seen in the upper panel).

\subsection{Present Day Properties and Predictions}
\label{sec:PDCV}

One of the very interesting and perhaps unique properties of KIC 5608384 concerns the mass of the WD accretor.  This is the lowest, robust inference of the mass of a WD accretor in a CV  ($0.46 \pm 0.02 M_\odot$) that has been reported to date.  The average mass of a CO WD in the field is $\simeq 0.6\,M_\odot$, which itself is considerably lower than the $0.83\,M_\odot$ average inferred for accreting WDs in CVs \citep{zorotovic11}.  This result is in stark contrast to the predicted mass distributions of standard CV evolution models (but see \citet{schreiberetal16}).  It should be noted that our result is based solely on the MC simulations using the observed properties and constraints (e.g., Roche geometry), and is not inferred using any information derived from the theoretical evolutionary tracks.  In fact, this inferred mass is close to the lower limit required by the theoretical models in order to avoid dynamical instabilities.  WD accretors with even lower masses would require lower-mass primordial secondaries (donors) and these could not reach a sufficiently evolved state in less than 10 Gyr. Not only do we conclude that the primordial conditions for KIC 5608384 place it close to the bifurcation boundary but that it is also close to the stability limit.

Assuming that our calculated value of the long-term secular average for the mass-transfer rate is correct, we can determine whether KIC 5608384 should experience disk instabilities (e.g., dwarf novae) or whether the disk should be in a high state (e.g., novalike CVs).  The transient nature of most CVs is likely due to a thermal-viscous disk instability (see, e.g., \citealt{Lin85}; \citealt{Cannizzo}; \citealt{Hameury}; \citealt{JPL}).  In CVs it is believed that the disk instabilities are dominated by the mass transfer rate.  The governing equation is Eqn (A.4) in \citep{JPL}  which derives the critical value of $\dot M$ at radius $r$ in the accretion disk.  After correcting for an error in the determination of the Roche-lobe radius (Eqn.~(34) in \citealt{JPL}), we find the following expression for the critical value of $\dot M$ required for stability:
\begin{equation}
\dot M_{\rm crit} \simeq 3.3 \times 10^{-9} f(q)^{2.68} (M_{\rm tot}/M_{\rm wd})^{0.89} P_{\rm hr}^{1.79}  \quad M_\odot ~{\rm yr}^{-1}
\end{equation}
where $f(q)$ is Eggleton's (1983) expression for the size of the Roche lobe, $q \equiv M_{\rm wd}/M_{\rm donor}$, $M_{\rm tot} \equiv M_{\rm wd}+M_{\rm donor}$, and $P_{\rm hr}$ is the orbital period in hours.{\footnote{Note that $M_2 \equiv M_{donor}$.}}  Using this equation, we find that $\dot M_{\rm crit} \simeq 1.2 \times 10^{-8} M_\odot ~{\rm yr}^{-1}$.  Given that our inferred value of the mass-transfer rate is $\sim 3 \times 10^{-10} M_{\odot}$/yr and that our theoretical, long-term secular average rate  is $\approx 6.5 \times 10^{-9} M_{\odot}$/yr, both of which are less than the critical value, we conclude that KIC 5608384 should exhibit dwarf novae phenomena.

As mentioned in subsection \ref{sec:evolve2}, the surface hydrogen abundance for the donor of KIC 5608384 should be close to the primordial value (and thus the predicted under-abundance should not be detectable).  However, it should be possible with HST to infer the (relative) abundances of at least two of the isotopes of $^{12}$C, $^{14}$N, or $^{16}$O with sufficient precision to be able to test our model.  Specifically, we predict that the (C/N) isotopic ratio should be a factor of 4 lower, the ratio of (O/N) should be a factor of 2.5 lower, and the (C/O) ratio should be 60\% of the primordial value. Ultraviolet spectroscopy of CVs with evolved donors often shows mildly to extremely depressed C/N ratios, e.g. \citet{gaensickeetal03}. While confirmation of these values would not necessarily validate our model, a contradiction of our predictions would certainly help to falsify it.

Finally, it is clear from the evolutionary tracks that KIC 5608384 will become an AM CVn-type system (see \citealt{Green}, and references therein).  These binaries are characterized by their ultra-short orbital periods ($5 \lesssim  P/{\rm min} \lesssim 50$), little or no detectable hydrogen in the matter accreted by the WD companion, and typically high mass-transfer rates (although this is mostly due to an observational selection effect).  Based on our evolutionary tracks we predict that KIC 5608384 will evolve to become an AM CVn in $\sim 1$ Gyr and that it will reach an absolute minimum period of $\simeq 20$ minutes at which point the donor's mass would be $\simeq 0.08 \, M_{\odot}$.  We would expect mass transfer to continue indefinitely until the donor has been whittled down to planetary masses.

\subsection{A Brief Look at the Properties of Long-Period CVs}

\begin{figure}
\begin{center}
\includegraphics[width=1.01 \columnwidth]{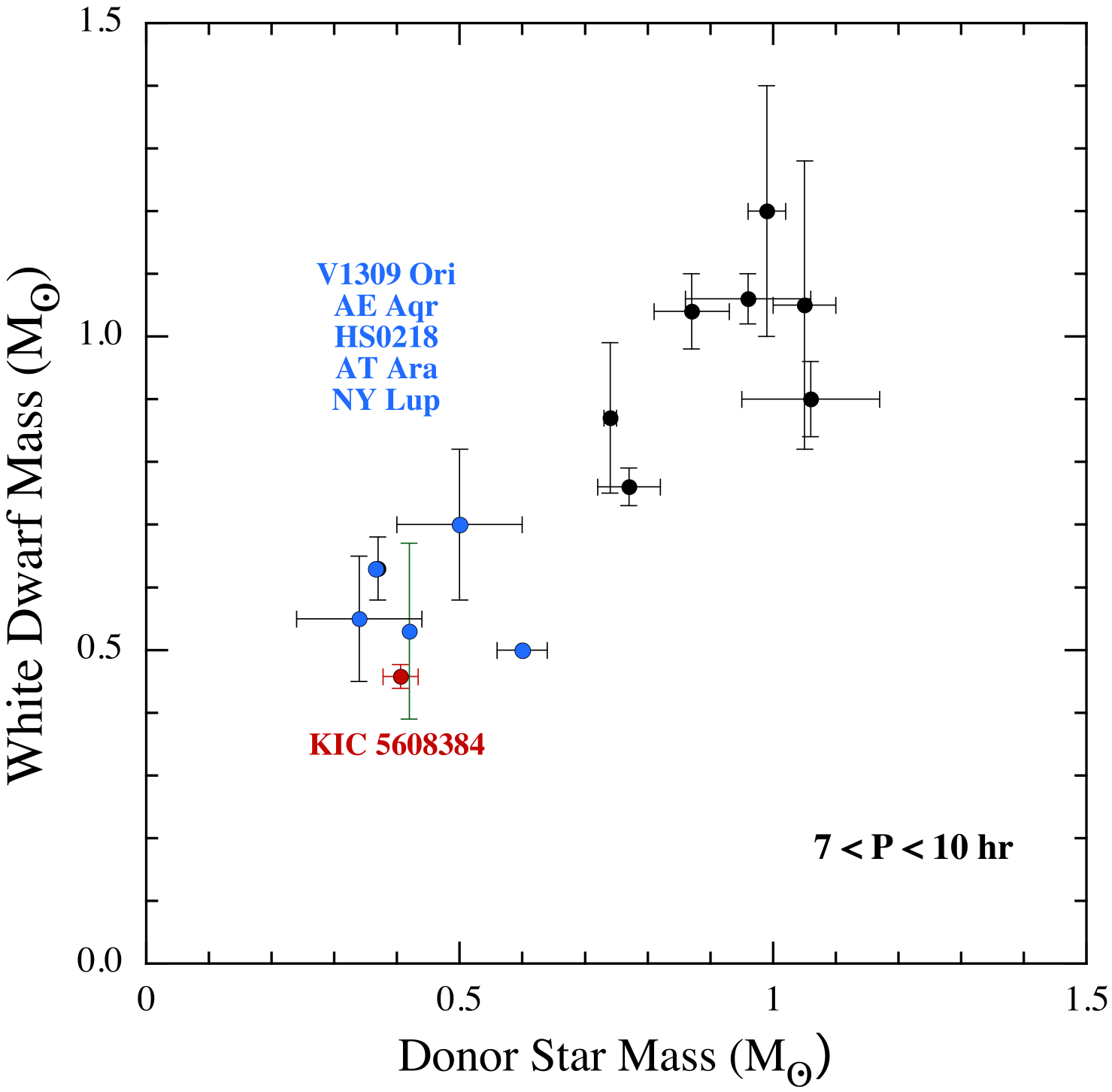}
\caption{White-dwarf mass vs.~donor-star mass for all CVs listed in the \citet{ritter03} catalog with measured masses and $7 < P_{\rm orb} < 10$ hr.  Possible close cousins to KIC 5608384 are marked in blue (V1309 Ori, AE Aqr, HS0218, AT Ara, and NY Lup; see Table \ref{tbl:cousins}). The label key is arranged in decreasing order of $M_{\rm wd}$.}
\label{fig:cousins} 
\end{center}
\end{figure}

Here we compare KIC 5608384 with some potential `cousins'.  In Fig.~\ref{fig:cousins} we show all the CVs from the \citet{ritter03} catalog with periods between 7 and 10 hours in the $M_{\rm wd}-M_{\rm  donor}$ plane.  We find five other systems that appear to be `cousins' of KIC 5608384 (marked in blue), albeit with somewhat larger uncertainties.  These all have WD masses in the range of $\simeq 0.5-0.7 \, M_\odot$ (Table\,\ref{tbl:cousins}), significantly below the average value of $\langle M_{\rm wd}\rangle=0.83\,M_\odot$ \citep{zorotovic11}~--~in fact, their white dwarfs are among the lowest-mass ones among all known CVs. Their donor stars have masses of $\simeq 0.35-0.6\,M_\odot$, which are much lower than the $\simeq1 M_\odot$ of a Roche-lobe filling main-sequence star at $P_\mathrm{orb}$.

Just as for KIC\,5608384, these five systems show signatures of having undergone significant nuclear evolution, probably have experienced thermal-timescale mass transfer at an earlier epoch, and are evolving to ultimately becoming AM\,CVn systems. AE\,Aqr has been discussed in detail by \citet{schenker02} as the archetypical post-thermal time-scale mass transfer CV; the donor stars in AT\,Ara, HS\,0218+3229, and NY\,Lup are significantly oversized for their mass \citep{bruch02, rodriguez-gil09, demartino06}, and V1309\,Ori shows the fingerprint of CNO processed material both in the ultraviolet via a large N/C emission line flux ratio \citep{szkody96} and in the
infrared via the absence of CO bands \citep{howell10}. 

We conclude that these likely progenitors of AM\,CVn systems make up a sizable fraction of the known CVs within the 7--10 hour period
range. Unfortunately, the general lack of good mass measurements for many of these systems makes it difficult to quantify this contribution.

\section{Summary} 
\label{sec:concl}

In this work, we have reported on the discovery of an unusual cataclysmic variable in the {\em Kepler} main field, KIC 5608384, which experienced only a single outburst over the four years of the observations (Sect.~\ref{sec:Kepler}).  The object was found via a visual inspection of all the {\em Kepler} main-field lightcurves. The quiescent data exhibit ellipsoidal light variations with a $\sim$12\% amplitude and period of 8.7 hours (see Sect.~\ref{sec:Kepler}).

\begin{table}
\centering
\caption{`Cousins' of KIC 5608384}
\begin{tabular}{lcccc}
\hline
\hline
CV & $M_{\rm wd}$ & $M_{\rm don}$ & Per. & Reference$^a$  \\
   &   $M_\odot$  &  $M_\odot$ & hr & \\
\hline
HS0218 & $0.55 \pm 0.10$ & $0.34 \pm 0.10$ & 7.13 & R09 \\   
V1309 Ori  & $0.70 \pm 0.12$ &  $0.50 \pm 0.10$  &  7.98 & G03 \\
AT Ara  & $0.53 \pm 0.14$ & $0.42$ & 9.01 &  B02 \\
NY Lup  &  $0.50$ &  $0.60 \pm 0.04$  &  9.86  & dM03  \\
AE Aqr  &  $0.63 \pm 0.05 $ &  $0.37$  &  9.88  & S02  \\
K5608384 & $0.46 \pm 0.02$ & $0.41 \pm 0.03$ & 8.74 & current work \\
\hline
\label{tbl:cousins}  
\end{tabular}

(a) References: R09 = \citet{rodriguez-gil09}; G03 = \citet{gaensicke03}; B02 = \citet{bruch02}; dM06 = \citet{demartino06}; S02 = \citet{schenker02}. For some of the mass values there are no available uncertainties. 
\end{table} 

We presented a series of ground-based spectra of KIC 5608384 from which we derived the radial velocity curve for the companion K star and the corresponding mass function (Sect.~\ref{sec:spectra}).  H$\alpha$ emission lines were present in the spectra even though these were  taken while the source was presumably in quiescence.  The H$\alpha$ lines are at least partially eclipsed by the companion K star.

The spectral energy distribution (`SED') for KIC 5608384, coupled with the Gaia distance,  was used to infer a radius of the K-star companion of $R_{\rm K} = 0.754 \pm 0.040\,R_\odot$ (Sect.~\ref{sec:SED}).  We then use the measured mass function, the amplitude of the ELVs, and the radius of the K star, in conjunction with an assumed Roche-lobe-filling geometry, to infer the system parameters (Sect.~\ref{sec:mcmc}).  We find the masses and orbital inclination angle to be: $M_{\rm wd} \simeq 0.46 \pm 0.02 \, M_\odot$, $M_{\rm K} \simeq 0.41 \pm 0.03 \, M_\odot$, and $i \gtrsim 70^\circ$, respectively.  Our estimate for the mass of the accreting WD is lower than any previously reported values for WDs in cataclysmic variables. 

We have also generated binary evolution tracks using {\tt MESA} to model the current status of KIC 5608384 as well as to understand its origins (Sect.~\ref{sec:origin}).  We believe that it is quite reasonable to believe that the primordial binary was formed about 8 to 9 Gyr ago and consisted of an $\approx 1.7 \, M_{\odot}$ primary and an  $\approx 1.1 \, M_{\odot}$ secondary in a 300-day orbit.  If correct, the core of the primary likely experienced a hot subdwarf phase after it lost its envelope due to common envelope ejection.  Based on the currently inferred properties of KIC 5608384 (especially the donor's large radius given its small mass), it is also likely that the ZACV started its evolution very close to the bifurcation limit (necessitating that the donor star be quite chemically evolved).  Based on this degree of nuclear evolution of the donor, we conclude that KIC 5608384 will likely evolve to become an AM CVn system in less than one billion years.  

\vspace{0.3cm}
\noindent
{\bf Acknowledgements}
We thank the referee, Linda Schmidtobreick, for constructive suggestions to improve the paper.
L.\,N.~thanks the Natural Sciences and Engineering Research Council (Canada) for financial support through the Discovery Grants program.  Some computations were carried out on the supercomputers managed by Calcul Qu\'ebec and Compute Canada. The operation of these supercomputers is funded by the Canada Foundation for Innovation (CFI), NanoQu\'ebec, R\'eseau de M\'edecine G\'en\'etique Appliqu\'ee, and the Fonds de recherche du Qu\'ebec -- Nature et technologies (FRQNT).  J.\,A.~ thanks NSERC (Canada) for an Undergraduate Student Research Award (USRA).  T.\,B. thanks the financial support of the (Hungarian) National Research, 
Developement and Innovation Office (NKFIH) through the grants OTKA K-113117 and NKFI KH-130372.  A.\,V.'s work was supported in part under a contract with the California Institute of Technology (Caltech)/Jet Propulsion Laboratory (JPL) funded by NASA through the Sagan Fellowship Program executed by the NASA Exoplanet Science Institute.  This paper includes data collected by the {\em Kepler} mission. Funding for the {\em Kepler} mission is provided by the National Aeronautics and Space Administration (NASA) Science Mission directorate.






\end{document}